\documentclass[journal]{IEEEtran}

\usepackage{amsmath}
\usepackage{amssymb}
\usepackage{amsthm}
\usepackage{graphicx}
\usepackage{subfigure}
\usepackage{multirow}
\usepackage{subfigure}
\usepackage{algorithm}
\usepackage{algorithmicx}
\usepackage{algpseudocode}
\usepackage{color}
\usepackage{threeparttable}
\usepackage{bm}
\usepackage{stfloats}
\usepackage{bigstrut}
\usepackage{booktabs}
\usepackage{soul}
\usepackage{caption}
\usepackage{cite}
\usepackage{flushend}
\usepackage[table]{xcolor}
\usepackage{array}
\usepackage[
colorlinks,
linkcolor=red,
anchorcolor=green,
filecolor=red,
urlcolor=red,
citecolor=green]{hyperref}
\usepackage[numbers,sort&compress]{natbib}
\soulregister\cite7
\soulregister\ref7

\newcommand{\RNum}[1]{\uppercase\expandafter{\romannumeral #1\relax}}

\begin{document}

	\title{Towards Top-Down Just Noticeable Difference Estimation of Natural Images}
	
	\author{Qiuping~Jiang,
		Zhentao~Liu,
		Shiqi~Wang,
		Feng~Shao,
		and Weisi~Lin,~\IEEEmembership{Fellow,~IEEE}
		\thanks{This work was supported in part by the Zhejiang Natural Science Foundation under Grant LR22F020002, in part by the Natural Science Foundation of China under Grants 61901236 and 62071261, and in part by the Fundamental Research Funds for the Provincial Universities of Zhejiang under Grant SJLZ2020003. \emph{(The first two authors contribute equally to this work. Corresponding author: Qiuping Jiang)}}
		\thanks{Q. Jiang, Z. Liu, and F. Shao are with the School of Information Science and Engineering, Ningbo University, Ningbo 315211, China (e-mail: jiangqiuping@nbu.edu.cn, zhentaoliu0319@163.com, shaofeng@nbu.edu.cn).}
		\thanks{S. Wang is with the Department of Computer Science, City University of Hong Kong, Kowloon Tong, Hong Kong (e-mail: shiqwang@cityu.edu.hk).}
		\thanks{W. Lin is with the School of Computer Science and Engineering, Nanyang Technological University, Singapore (e-mail: wslin@ntu.edu.sg).}
	}
	
	\markboth{IEEE Transactions on Image Processing}
	{Shell \MakeLowercase{\textit{et al.}}: Bare Demo of IEEEtran.cls for IEEE Journals}
	
	\maketitle
	
	\begin{abstract}
		Just noticeable difference (JND) of natural images refers to the maximum pixel intensity change magnitude that typical human visual system (HVS) cannot perceive. Existing efforts on JND estimation mainly dedicate to modeling the diverse masking effects in either/both spatial or/and frequency domains, and then fusing them into an overall JND estimate. In this work, we turn to a dramatically different way to address this problem with a top-down design philosophy. Instead of explicitly formulating and fusing different masking effects in a bottom-up way, the proposed JND estimation model dedicates to first predicting a critical perceptual lossless (CPL) counterpart of the original image and then calculating the difference map between the original image and the predicted CPL image as the JND map. We conduct subjective experiments to determine the critical points of 500 images and find that the distribution of cumulative normalized KLT coefficient energy values over all 500 images at these critical points can be well characterized by a Weibull distribution. Given a testing image, its corresponding critical point is determined by a simple weighted average scheme where the weights are determined by a fitted Weibull distribution function. The performance of the proposed JND model is evaluated explicitly with direct JND prediction and implicitly with two applications including JND-guided noise injection and JND-guided image compression. Experimental results have demonstrated that our proposed JND model can achieve better performance than several latest JND models. In addition, we also compare the proposed JND model with existing visual difference predicator (VDP) metrics in terms of the capability in distortion detection and discrimination. The results indicate that our JND model also has a good performance in this task. The code of this work are available at \url{https://github.com/Zhentao-Liu/KLT-JND}.
	\end{abstract}
	
	
	\IEEEpeerreviewmaketitle
	
	\section{Introduction}
	Just noticeable difference (JND) refers to the maximum pixel intensity change magnitude that typical human visual system (HVS) cannot perceive \cite{JNDSurvey1}. It reveals the visibility limitation of the HVS and reflects the underlying perceptual redundancy in visual signals, rendering it useful in many perceptual image/video processing applications such as image/video compression \cite{PVSCT,JNDPOJPEG,ECRJND,YANG}, perceptual image/video enhancement \cite{SESIUBJND,PIESI}, information hiding and watermarking \cite{PTWSCI,JNDGIW,JNDWM,JNDMAWDW,TDIHODS,EPSEODEI,SLSB}, and visual quality assessment \cite{PVQM,SIQM,AWTS,RRIQA,JNDIQA}, etc. Due to its wide applications, JND estimation of natural images has received much attention and been widely investigated. Generally, the JND models can be divided into two categories: JND models in pixel domain and JND models in frequency domain.
	
	JND models in pixel domain directly calculate the JND at pixel level with considerations of either/both luminance adaption (LA) or/and contrast masking (CM) effects of the HVS. As the pioneering work, Chou \textit{et al.} \cite{subJND} first proposed a spatial-domain JND model by combining LA and CM. Afterwards, Yang \textit{et al.} \cite{Yang2005} further proposed a generalized spatial JND model with a nonlinear additivity model for masking effects (NAMM) to characterize the possible overlaps between LA and CM. Based on Yang \textit{et al}.'s work, Liu \textit{et al.} \cite{JNDDMSETR} introduced an enhanced pixel-level JND model with an improved scheme for CM estimation. Specifically, the image is first decomposed into structural component (\textit{i.e.}, cartoon like, piecewise smooth regions with sharp edges) and textural component using a total-variation algorithm. Then, the structural and textural components are used estimating EM and TM effects, respectively. In order to differently manipulate order and disorder regions, Wu \textit{et al.} \cite{Wu2013} designed a novel JND estimation model based on the free-energy principle. An autoregressive model is first applied to predict the order and disorder contents of an input image, then different schemes are used to estimate the JND threshold of these two parts, respectively. Further, Wu \textit{et al.} \cite{Wu2017} took the concept of pattern complexity (PC) into account for JND estimation. They quantified the pattern complexity as the diversity of pixel orientations. Finally, pattern masking is deduced and combined with the traditional CM for JND estimation. Jakhetiya \textit{et al.} \cite{JAKHETIYA} further combines root mean square (RMS) contrast with LA and CM to build a more comprehensive JND model in low-frequency regions. For high-frequency regions, a feedback mechanism is used to efficiently mitigate the over- and under-estimations of CM. Chen \textit{et al.} \cite{Chen2020} took horizontal-vertical anisotropy and vertical-meridian asymmetry into consideration to yield a better JND estimation. The basic consideration is that the effect of eccentricity on visual sensitivity is not homogeneous across the visual field. Shen \textit{et al.} \cite{Shen2021} decompose the image into three components namely luminance, contrast, and structure. Since the masking of structure visibility (SM) is unknown, they trained a deep learning-based structural degradation estimation model to approximate SM. Finally, LA, CM and SM are combined to estimate the overall JND. Wang \textit{et al.} \cite{Wang2021} proposed a novel JND estimation model by exploiting the hierarchical predictive coding theory. They simulated both the positive and negative perception effects of each stage individually and integrated them with Yang \textit{et al.}'s NAMM model \cite{Yang2005} to get the final JND.
	
	JND models in frequency domain firstly transform the original image into a specific transform domain and then the corresponding JND thresholds for each sub-band are estimated. The main consideration of these frequency domain-based JND models is to make use of the well-known contrast sensitivity function (CSF) which reflects the bandpass characteristics of HVS in the spatial frequency domain and is typically modeled as an exponential function of the spatial contrast \cite{JNDsurvey}. In \cite{VideoJND}, the visibility thresholds for different frequencies are measured through subjective tests, and the CSF is built to account for the fundamental/base JND threshold for each sub-band. Typically, the JND thresholds for each sub-band are usually estimated based on a fixed size block (e.g., $8\times8$) via a linear combination of CSF and some other modulation factors. For example, Wei and Ngan \cite{STJND} utilized a simple piecewise function to represent LA and formulated the CM as a categorizing function according to the richness of block texture information. Bae \textit{et al.} \cite{BlockJND} proposed a new DCT-based JND profile by incorporating the CSF, LA, and CM effects. Specifically, a new CM JND is modeled as a function of DCT frequency and a newly proposed structural contrast index (a new texture complexity metric that considers not only contrast intensity, but also structureness of image patterns). Wan \textit{et al.} \cite{JNDORDCT} analyzed orientation information with the DCT coefficients and a more accurate CM model was proposed in the DCT domain. Besides LA and CM, some works also considered foveated masking as influential factors for JND estimation. The foveated masking was first modeled by Chen \textit{et al.} in \cite{Chen2020} where it was utilized as an explicit form in pixel-domain JND and then incorporated in the frequency-domain JND estimation.
	
	The above only provides a brief overview of exsiting advances in the field of JND estimation and a more comprehensive survey can be found in \cite{JNDSurvey1,JNDsurvey}. In general, the existing JND models share the same design philosophy, i.e., explicitly modeling diverse visibility masking effects and then fusing them together to derive an overall JND estimation. Such a kind of design philosophy can be considered as a bottom-up strategy which starts from all possible concrete influential factors and then progressively produce the final estimation. We claim that this design philosophy has some inherent drawbacks. First, without having a deep understanding of the HVS properties, it is hard to take all visibility masking effects into account. Second, the interactions among different masking effects are difficult to be characterized. Therefore, the performance of the existing JND estimation models remains limited. 
	
\begin{figure}[!t]
		\centering
		\includegraphics[width=0.48\textwidth]{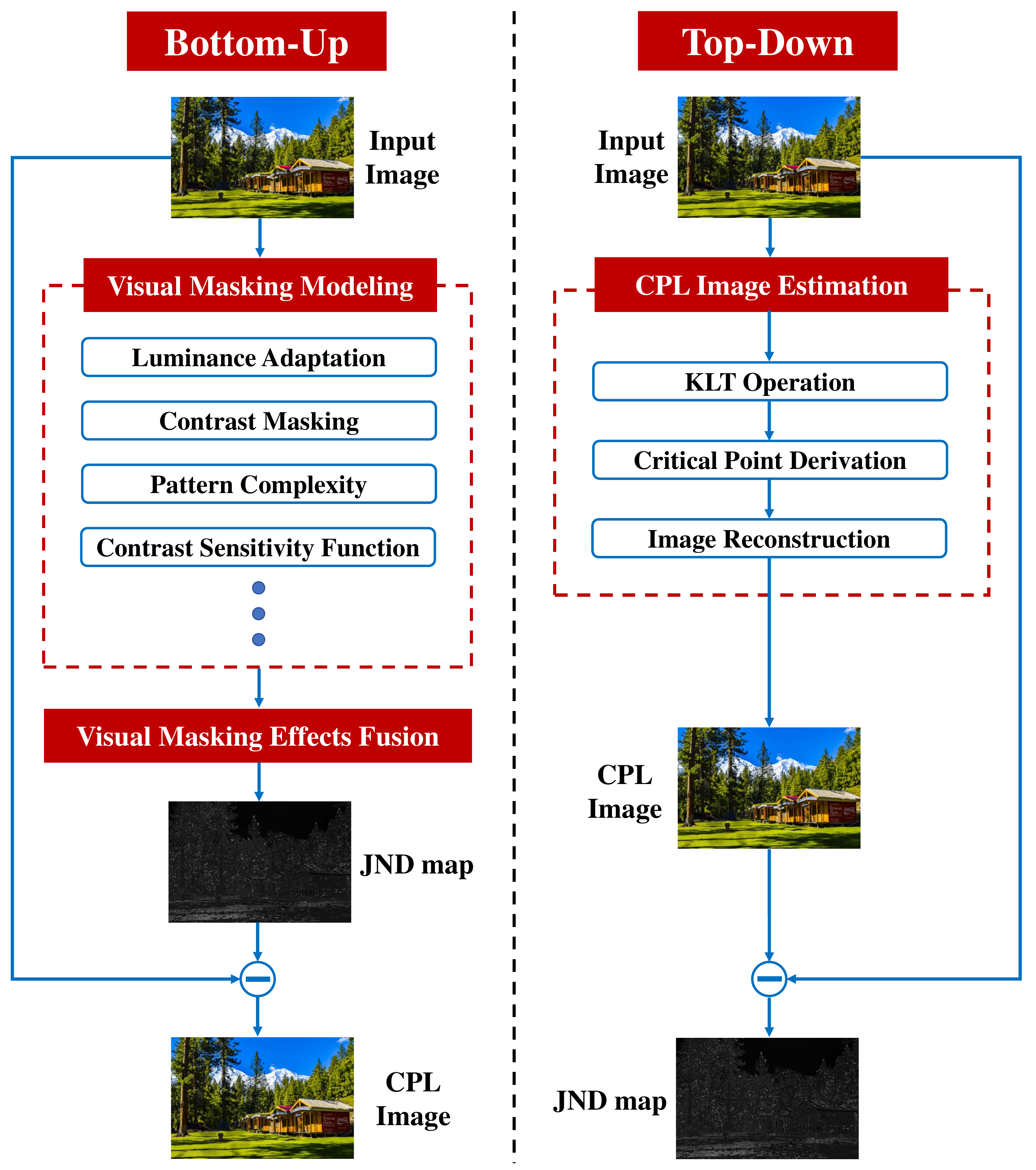}
		\caption{Pipelines of bottom-up and top-down design philosophies for JND estimation.}
		\label{fig_pipeline}
	\end{figure}

	In this paper, we turn to a different way to address these problems from a top-down perspective. Keeping in mind that the goal is to estimate the maximum change magnitude (JND threshold) of each pixel that typical HVS cannot perceive. In other words, with an ideal JND map, if the original image is changed within the JND threshold, we can still perceive the changed image as a perceptual lossless one. Thus, we intuitively come up with the idea to firstly determine a critical perceptual lossless (CPL) image and then calculate the difference between the CPL image and the original image as the JND map, without explicit modelling and fusion of different visibility masking effects. The CPL image refers to a changed image whose change magnitude is just at the critical point beyond which this changed image will be either perceived as lossy or not the critical one (although perceptual lossless). To facilitate comparing the bottom-up and top-down design philosophies for JND estimation, we compare their pipelines in Fig. \ref{fig_pipeline}. As shown, the top-down JND estimation (right column) first starts from the prediction of a CPL image from the original one and then produces the final JND map by calculating their difference. However, the traditional bottom-up JND estimation (left column) first starts from the explicit modelling and fusion of different visibility masking effects to derive the overall JND map based on which a perceptual lossless image is expected to be obtained (the final result may not be perceptual lossless since the JND map established in this way may not be sufficiently accurate). 
	
	Given an input image, we first perform the Karhunen-Lo\'{e}ve Transform (KLT) and then derive its critical point (i.e., perceptual lossless threshold) by exploiting the convergence characteristics of KLT coefficient energy. Once the critical point is determined, the CPL image can be reconstructed. Then, the difference map between the original image and the CPL image is deemed as the visually redundant information that cannot be perceived by HVS, implying the visibility limitation of the HVS. Finally, we just simply take the derived difference map as the final JND map. 
	
	Although we have mentioned that one of the main shortcomings of the current bottom-up JND estimation approaches is the insufficient understanding of the HVS properties, we do not intend to interpret all perceptual processes with KLT. Instead, we just try to address the JND estimation problem from another perspective by exploiting certain properties of KLT to identify the critical point (indicating how many spectral components are involved) for reconstructing a CPL image. Although the objective is to estimate JND map, the starting points of the conventional JND models and our proposed JND model are different. That is, the conventional JND models usually follow a bottom-up design philosophy which starts from modeling the visibility masking effects of different factors and then fusing them together to obtain a final JND estimation, while the proposed JND model is designed from another perspective, i.e., top-down perspective, which dedicated to determining the CPL as accurate as possible. Obviously, the modeling of masking effect of different influential factors requires substantial knowledge about the HVS properties while the determination of CPL is much simpler in this sense. In the meanwhile, we notice that there are some studies on investigating the visually lossless threshold of image compression \cite{viewing-condition1,viewing-condition2}, which is similar to the critical point defined in our paper. However, compared with the widely used quality factor which is mainly applicable to image compression \cite{jin2016statistical,viewing-condition1,viewing-condition2}, the spectral component in KLT is not restricted to any specific applications. In the literature, there is also another concept, i.e., Visual Difference Predictor (VDP) \cite{HDR-VDP}, which is a relevant concept with JND. However, their definitions are different. The JND model and VDP metric have different input signals and their outputs also have different physical meanings. Let us denote the original image as $I_o$, the distorted image as $I_d$, the JND model as $F_{JND}$, the output of the JND model as $O_{JND}$, the VDP metric as $F_{VDP}$, and the output of the VDP metric as $O_{VDP}$. The mathematical definitions of a JND model and a VDP metric can be briefly expressed as follows:
	\begin{equation}
		O_{JND} = F_{JND}(I_o)
	\end{equation}
	\begin{equation}
		O_{VDP} = F_{VDP}(I_o,I_d)
	\end{equation}
	According to the above mathematical formulations, we observe that the input of a JND model is the original image $I_o$ and the output of a JND model is the JND map $O_{JND}$. The value of each pixel in the JND map $O_{JND}$ represents the maximum tolerant intensity value variation of each pixel in the original image $I_o$. Different from the JND model, a VDP metric takes the original image $I_o$ and the distorted image $I_d$ as inputs. The output of a VDP metric is a probability map $O_{VDP}$. The value of each pixel in $O_{VDP}$ represents the probability of detecting distortion at the corresponding pixel in the distorted image $I_d$. Overall, JND is a computational model of visual redundancy measured on a single original image while VDP is a distortion visibility metric measured on an image pair including an original image and its distorted counterpart. Although conceptually different, these two concepts are still relevant because they both attempt to model the same underlying mechanism of the visual system - detection and discrimination. In Section III-D, we will show how the JND map can be converted into a VDP-type metric and directly compared with existing VDP metrics.
	
	The contributions of this work are as follows: 1) We propose a novel JND model for natural images from a top-down perspective without explicitly modeling and fusing different masking effects; 2) We transfer the problem of JND estimation into deriving a CPL counterpart of its original image and resort to the KLT theory for the first time to derive the critical point; 3) We conduct subjective experiments to determine the critical points of 500 images and find that the distribution of cumulative normalized KLT coefficient energy values at the critical points can be well characterized by a Weibull distribution. 
	The reminder of this paper is outlined as follows. Section II illustrates the proposed JND model including motivation and algorithm details. Section III presents the experimental results and comparisons with existing JND models. Finally, conclusions are drawn in section IV.

	\section{Methodology}
	\subsection{Design Philosophy and Motivation}
	As stated, we turn to a top-down design philosophy for JND estimation. As illustrated in Fig. \ref{fig_pipeline}, most of the traditional JND profiles follow a bottom-up design philosophy. Given an input image, they start from visibility masking effect modeling of multiple factors including LA, CM, PC, CSF, etc. Afterwards, those masking effects are fused together to obtain a final JND estimation via linear or nonlinear combinations. However, this bottom-up design philosophy suffers from some inherent drawbacks. First, the overall visibility masking effect of the HVS can be related with more contributing factors beyond those have been considered in the existing works and it is also insufficiently accurate to formulate the masking effect even for a single specific contributing factor. Moreover, the used linear or nonlinear models for different masking effect fusion are also unable to characterize the complex interactions among different masking effects. To overcome such drawbacks, we propose a top-down design philosophy. 
	
	Considering the visibility limitation of HVS, it is believed that a CPL counterpart of the input image exists. The CPL image refers to a changed image whose change magnitude is just at the critical point beyond which this changed image will be either perceived as lossy or not the critical one (although perceptual lossless). Since we just cannot perceive the difference in the CPL, thus the difference map between the original image and the CPL image is just the JND map according to the definition of JND. Thus, the problem of JND estimation can be transfered into a CPL image estimation problem. Compared with directly estimating the JND map, it is much easier and intuitive to derive a CPL image. In this work, we resort to the KLT to obtain the CPL image.
	
	KLT is a signal-dependent linear transform. As a data-driven transform, it KLT has been applied in image coding \cite{KLTcode} and image quality assessment \cite{MsKLT,KLTCIQA} with promising performance due to its excellent decorrelated performance. The KLT domain and spatial domain can be converted from one to another without loss of information. Given an input image, we first transfer it from the spatial domain to the KLT domain. The KLT coefficients associated with different spectral components are responsible for visual information in different aspects. The former spectral components are responsible for macro-structures image information while the latter spectral components  are responsible for micro-structures image information. To validate this point, Fig. \ref{fig_SCs} illustrates the image reconstruction process via inverse KLT, i.e., from KLT domain to spatial domain, with different numbers of spectral components. In this example, we set the total number of spectral components $K=64$. Fig. \ref{fig_SCs}(a) is the original image, Fig. \ref{fig_SCs}(b)-(h) are the reconstructed images with the first $k$ spectral components, where $k\in\{1,2,4,8,16,32,64\}$. As shown in Fig. \ref{fig_SCs}(b), when only the 1-st spectral component is involved for reconstruction, almost all the macro-structures are recovered. However, many small textures and fine details are missing. As $k$ increases, the small textures and fine details become richer and clearer. These observations well validate our previous statement that different spectral components are responsible for image information in different aspects. Generally, we have the insight that the front part of spectral components in the KLT coefficient matrix take charge of the reconstruction of image macro-structures such as the basic contour and main structures while the latter part of spectral components take charge of the reconstruction of image micro-structures such as the textures and fine details.
	
	Obviously, there is a redundancy of visual information in an image. As shown in Fig. \ref{fig_SCs}(f), when we take the first 16 spectral components to perform reconstruction via inverse KLT, the visual quality of the reconstructed image is almost the same with that of the original image, i.e., Fig. \ref{fig_SCs}(a). The image reconstruction process via inverse KLT demonstrates that some micro-structure image information is perceptually redundant to HVS and directly discard the corresponding spectral components in the KLT coefficient matrix for reconstruction would not cause visible visual quality degradation of the reconstructed image. We wonder that if there exists a ciritical point (perceptual lossless threshold) $L$ which satisfies the following property: when we take the first $L$ spectral components to perform image reconstruction via inverse KLT, all the macro-structure image information and sufficient micro-structure image information are recovered so as to reconstruct the CPL image. Obviously, the accurate determination of the critical point $L$ is a key step to the success of our JND model as different critical points $L$ will yield different CPL images as well as different JND maps.

	
	\subsection{JND Estimation Based on CPL Image Prediction}
	\subsubsection{Image Transfom Using KLT}
	KLT is a signal dependent linear transform, the kernels of
	which are derived by computing the principal components
	along eigen-directions of the autocorrelation matrix of the
	input data. Given an image $\mathbf{X}$ with size $M \times N$, a set of non-overlapping patches with size $\sqrt{K} \times \sqrt{K}$ are extracted. These image patches are vectorized and combined together to form a new matrix $\mathbf{X} = [\mathbf{x}_{1},\mathbf{x}_{2},\cdots,\mathbf{x}_{S}] \in  \mathbb{R}^{K \times S}$, where $\mathbf{x}_{s} \in \mathbb{R}^{K\times1}, s=1,2,\cdots,S$ represents the $s$-th vectorized patch and $S$ is the total number of image patches in $\mathbf{X}$. The covariance matrix of $\mathbf{X}$ is defined as follows
	\begin{align}
		\mathbf{C} &= \mathbb{E}[(\mathbf{x}_{s}-\mathbf{\bar{x}})(\mathbf{x}_{s}-\mathbf{\bar{x}})^{\mathrm{T}}] \\&=\frac {1}{S-1} \sum_{s=1}^{S}(\mathbf{x}_{s}-\mathbf{\bar{x}})(\mathbf{x}_{s}-\mathbf{\bar{x}})^{\mathrm{T}}
	\end{align}
	where $\mathbf{\bar{x}}=\frac {1}{S}\sum_{s=1}^{S}\mathbf{x}_{s}$ denotes the mean vector obtained by averaging each row of $\mathbf{X}$ and $\mathbf{C} \in \mathbb{R}^{K \times K}$. Then, the eigenvalues and eigenvectors of $\mathbf{C}$ are calculated via eigenvalue decomposition. The eigenvectors are arranged according to their corresponding eigenvalues in the descending order to form the KLT kernel $\mathbf{P}=[\mathbf{p}_{1},\mathbf{p}_{2},\cdots,\mathbf{p}_{K}] \in  \mathbb{R}^{K \times K}$ where $\mathbf{p}_{k} \in \mathbb{R}^{K\times1},\;k=1,2,\cdots,K$ represents the $k$-th eigenvector. Using the KLT kernel $\mathbf{P}$, the KLT of $\mathbf{X}$ is expressed as follows:
	\begin{equation}
		\mathbf{Y} = {\mathbf{P}}^{\mathrm{T}} \mathbf{X}
	\end{equation}
	where $\mathbf{Y} = [\mathbf{y}_{1},\mathbf{y}_{2},\cdots,\mathbf{y}_{K}]^{\mathrm{T}} \in \mathbb{R}^{K \times S}$ is the KLT coefficient matrix and $\mathbf{y}_{k} \in \mathbb{R}^{S\times1},\;k=1,2,\cdots,K$ refers to the $k$-th spectral component obtained by $\mathbf{y}_{k} = \left({\mathbf{p}_{k}}\right)^{\mathrm{T}}\mathbf{X}$.
	
	\begin{figure}[!t]
		\centering
		\includegraphics[width=\linewidth]{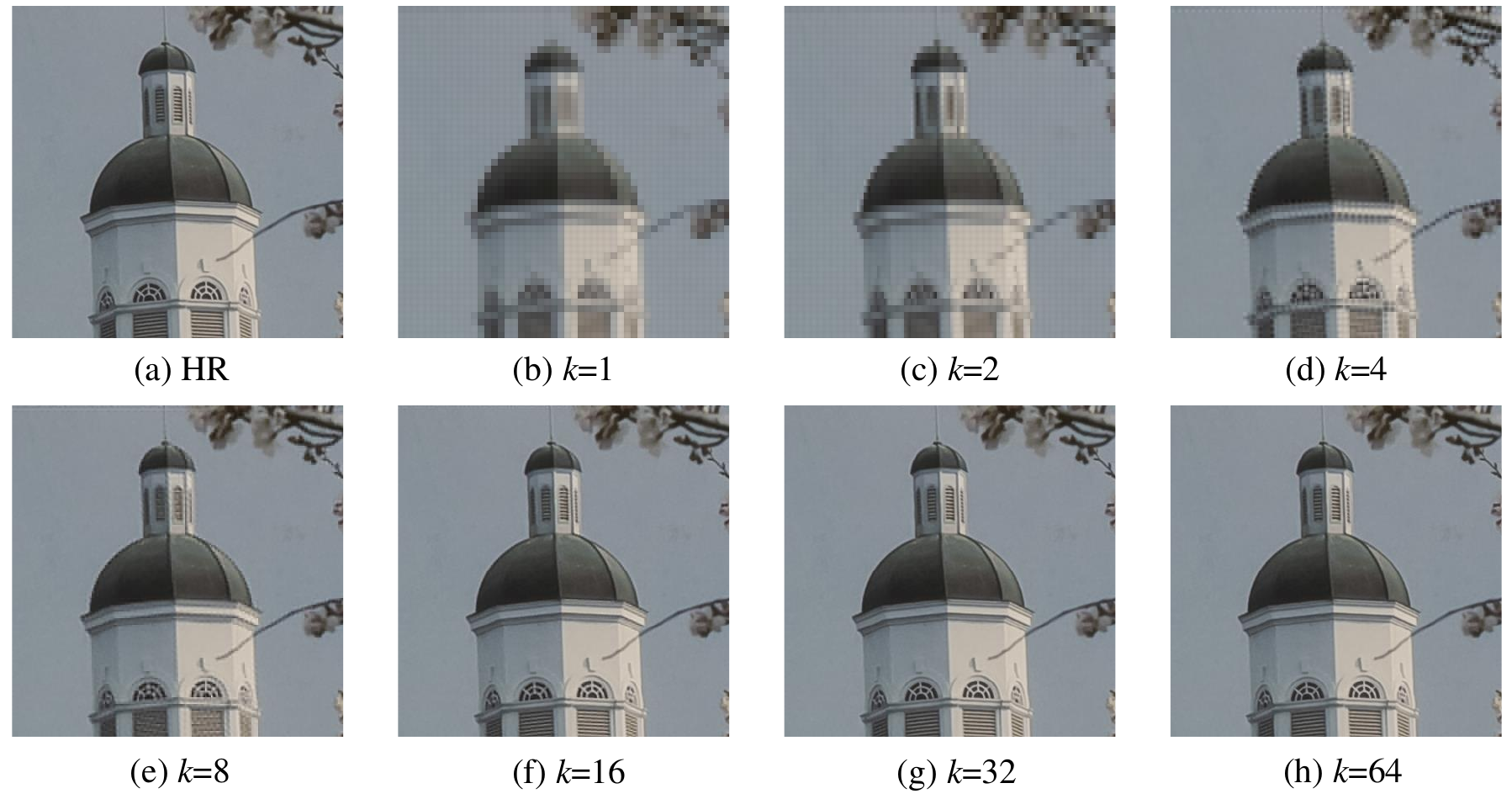}
		\caption{Reconstructed images with different numbers of spectral components in KLT. Zoon-in for best viewing.}
		\label{fig_SCs}
	\end{figure}
	
	
	\begin{figure*}[!t]
		\centering
		\includegraphics[width=\linewidth]{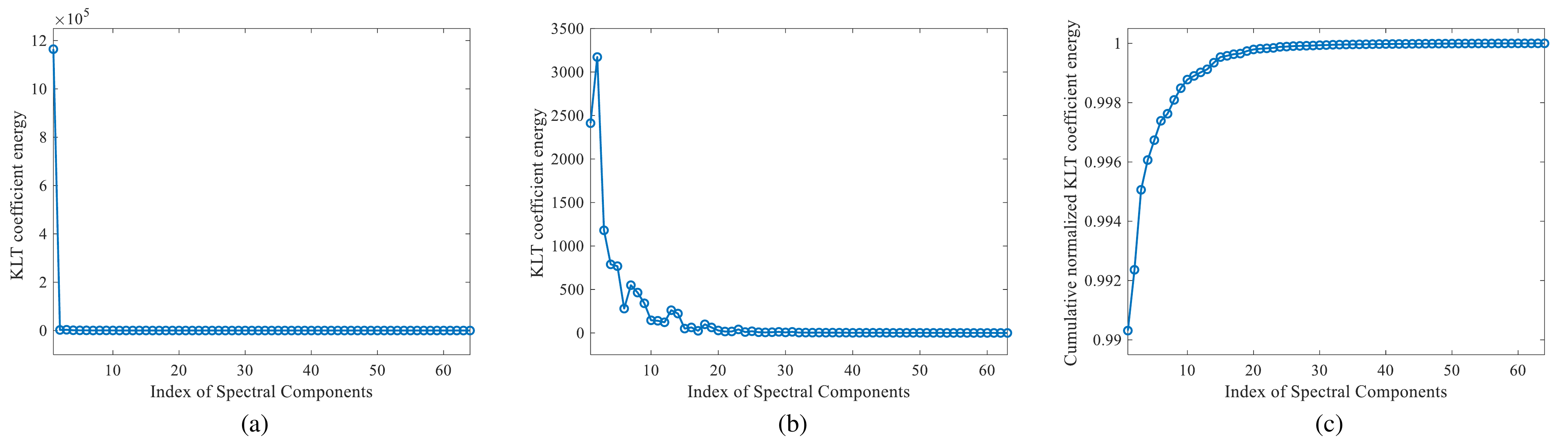}
		\caption{Distribution curves of the normalized KLT coefficient energy and cumulative normalized KLT coefficient energy.}
		\label{fig_KLTcoef}
	\end{figure*}

	\subsubsection{Convergence Property of KLT Coefficient Energy}
	After obtaining the KLT coefficient matrix $\mathbf{Y}$, we calculate the KLT coefficient energy $E_k$ for spectral component $y_k$ as follows: 
	\begin{equation}
		E_k=\frac{1}{S}\sum_{s=1}^{S}\mathbf{Y}(k,s)^2,\;\; k=1,2,\cdots,K.
	\end{equation}
	In order to remove the influence of image content, we further calculate the normalized KLT coefficient energy $p_k$ as follows:
	\begin{equation}
		p_k=\frac{E_k}{E_1+E_2+\cdots+E_K},\;\; k=1,2,\cdots,K.
	\end{equation}
	Fig. \ref{fig_KLTcoef}(a) shows the normalized KLT coefficient energy distribution curve for the image in Fig. \ref{fig_SCs}(a). Since $p_1$ is particularly large than others, we further plot another curve in Fig. \ref{fig_KLTcoef}(b) by excluding $p_1$. As shown in Fig. \ref{fig_KLTcoef}(b), as the spectral component index $k$ increases, the KLT coefficient energy $p_k$ first drops dramatically and later converges to be stable. When $k$ is larger than 20, $p_k$ becomes extremely small and gradually converges to zero. This phenomenon is consistent with the inverse KLT-based image reconstruction process illustrated in Fig. \ref{fig_SCs}. The former spectral components occupy much larger energies than the later ones so that they take charge of the macro-structure image information. The latter spectral components own relatively small energies such that they are in charge of the micro-structure image information.
	
	Now, let us take a look at the normalized KLT coefficient energy curve from an cumulative perspective. The cumulative normalized KLT coefficient energy  $P_k$ is obtained as follows:
	\begin{equation}
		P_k=p_1+p_2+\cdots+p_k,\;\; k=1,2,\cdots,K.
	\end{equation}
	The cumulative distribution curve of the normalized KLT coefficient energy for the image in Fig. \ref{fig_SCs}(a) has been shown in Fig. \ref{fig_KLTcoef}(c). Note that $P_1=p_1$ and has been already close to 1 due to the particularly large energy of the first spectral component. As $k$ increases, $P_k$ monotonously increases. When $k>20$, $P_k$ will gradually converges to 1, indicating the recovered visual information gradually become saturated. 
	
	\begin{figure}[!t]
		\centering
		\includegraphics[width=0.9\linewidth]{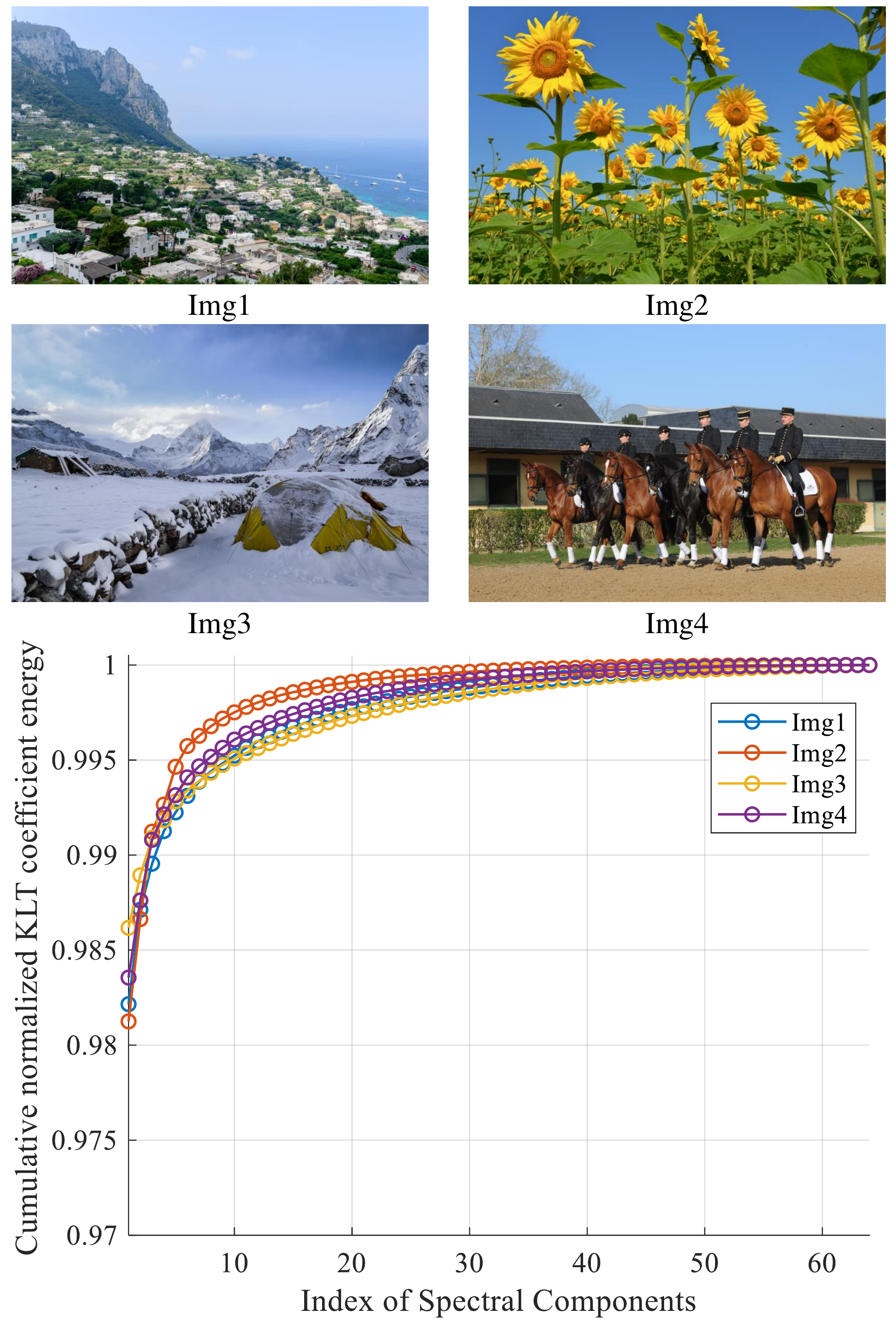}
		\caption{Cumulative distribution curve of the normalized KLT coefficient energy.}
		\label{fig_Cumulative}
	\end{figure}
	
	\subsubsection{Critical Point Derivation}
	For images with different contents, although their cumulative distribution curves have the same convergence tendency, their critical points that lead to sufficient visual information may be different, as demonstrated in Fig. \ref{fig_Cumulative}. Therefore, it is required to design an adaptive scheme to automatically determine the critical point for different images. In the following, we will illustrate how this issue is addressed with subjective studies and statistical analyses.
	
\begin{figure}[!t]
		\centering
		\includegraphics[width=0.95\linewidth]{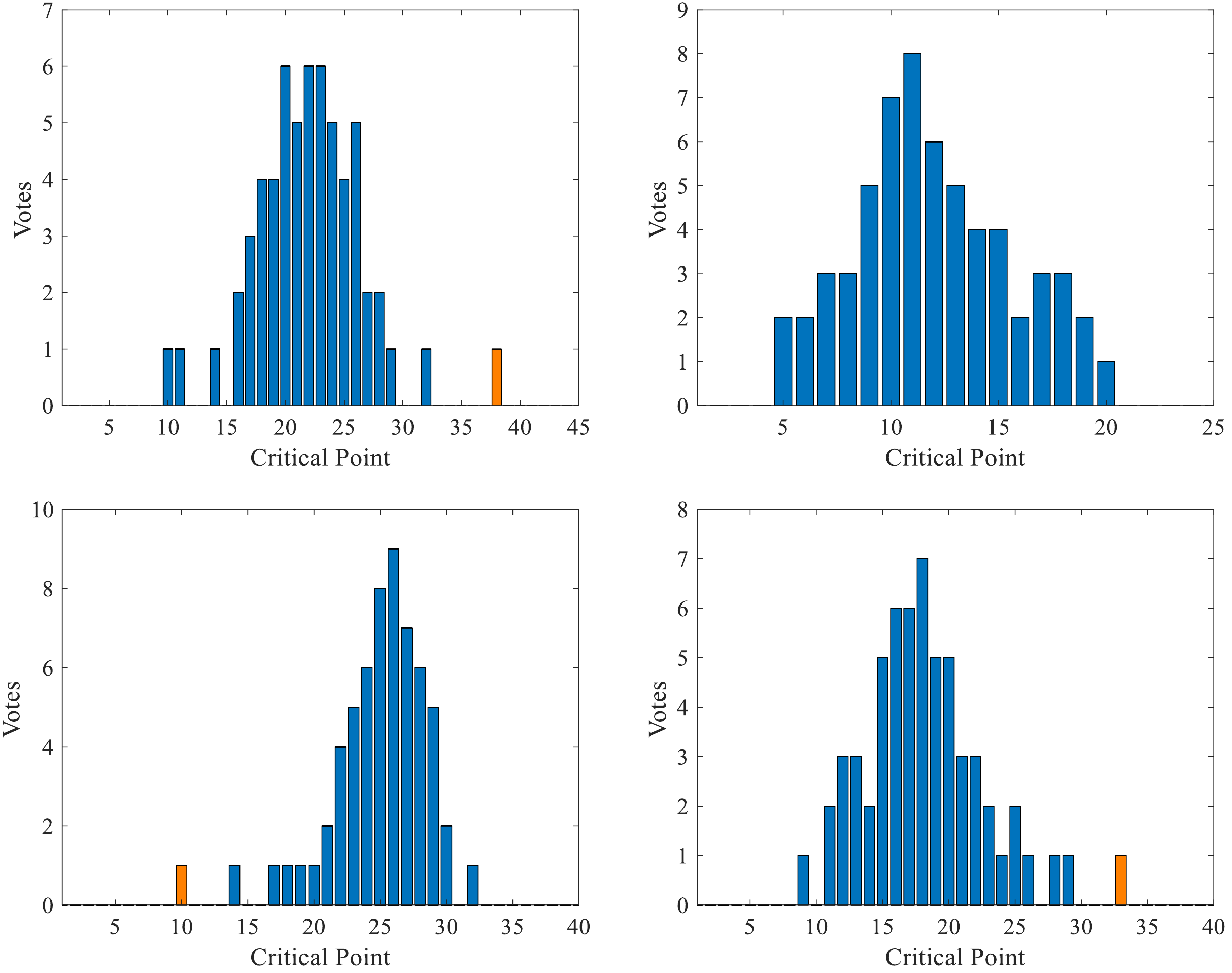}
		\caption{Visualization of vote distributions from all participants.}
		\label{fig_votedistribution}
	\end{figure}

	We select a total number of 500 high visual quality images from the DIV2K \cite{DIV2K1,DIV2K2} dataset. For each image, we apply the KLT-based image transform and inverse KLT-based image reconstruction with the first $k$ spectral components, $k=1,2,\cdots,K$. Thus, we can obtain $K$ reconstructed images for each original image. Then, we conduct a user study to subjectively determine the critical point $L\in\{1,2,\cdots,K\}$ for each original image. Details of the user study are illustrated as follows. There are 60 participants (38 males and 22 females) in our subjective experiment. Each participant $s^m$ is asked to compare each original image $I_o$ and its corresponding $K$ reconstructed versions $\{I_r^k\}$ one-by-one. During the subjective experiments, the image pairs are presented in a pre-defined order: $\{I_o,I_r^1\}$,$\{I_o,I_r^2\}$,$\cdots$,$\{I_o,I_r^K\}$. The image pairs are displayed on a DELL U2419HS monitor without resizing. The display is set to the sRGB color profile, the size of display is 23.8 inches (593.5mm$\times$353.5mm), the resolution is 1920$\times$1080, the peak luminance is set to 220$\rm cd/m^2$, the contrast is 1000:1, the viewing distance is 0.45m, and the pixels per visual degree (ppd) is 29.6266, approximately 30. All the subjective experiments are conducted in a dark room with dimmed lights. Each participant observes the presented image pairs carefully and conduct a binary judgment to answer the question whether the presented image pair has visible difference or not. Typically, the participants will observe obvious difference for the first several pairs and then gradually have difficulties in differentiating the later ones. For each participant, the user study will stop once he/she cannot observe any visible difference for a certain image pair $\{I_o,I_r^k\}$ and we cosider this spectral component index $k$ as the critical point given by the $m$-th participant $s^m$ for the $n$-th original image $I_o^n$: $L_{n}^{m}=k$. The experiment then continues untill each participant have finished determining the critical points for all 500 images. To understand the distribution of subjective votes, we also take the four images shown in Fig. \ref{fig_Cumulative} as examples and plot their vote distributions from all participants in Fig. \ref{fig_votedistribution}. We can see that the votes of critical points from all participants for each individual image approximate a Gaussian distribution. 
	
	\begin{table*}[!t]
		\centering
		\caption{Data analysis on vote distribution of different images.}
		\label{tab_dataanalysis}
		\setlength{\tabcolsep}{5pt}
		\renewcommand\arraystretch{1.5}
		\resizebox{430pt}{!}{\begin{tabular}{c|c|c|c|c|c|c|c}
				\hline \hline
				\multirow{1}{*}{Image} 
				& P-value & $\mu_{\mathbf{L}_{n}}$ & $\sigma_{\mathbf{L}_{n}}$ & $[\mu_{\mathbf{L}_{n}}-3\sigma_{\mathbf{L}_{n}},\mu_{\mathbf{L}_{n}}+3\sigma_{\mathbf{L}_{n}}]$ & $\mu_{\mathbf{L}^{'}_{n}}$ & $L_{n}$ &  $P_{L_{n}}$  \\  \hline \hline
				Img1 &0.11917318 &22.0000 &4.6080 &[8.1761, 35.8239] &21.7288 &22 &0.99787338  \\ \hline
				Img2 &0.28491706 &11.9667 &3.6695 &[0.9580, 22.9753] &11.9667 &12 &0.99802384  \\ \hline
				Img3 &0.00020296 &24.8361 &3.8033 &[13.4262,36.2459] &25.0833 &26 &0.99796027  \\ \hline
				Img4 &0.09625943 &18.2167 &4.5829 &[4.4679, 31.9654] &17.9661 &18 &0.99788544  \\ \hline \hline
		\end{tabular}}
	\end{table*}
	
	\begin{figure}[!t]
		\centering
		\includegraphics[width=0.95\linewidth]{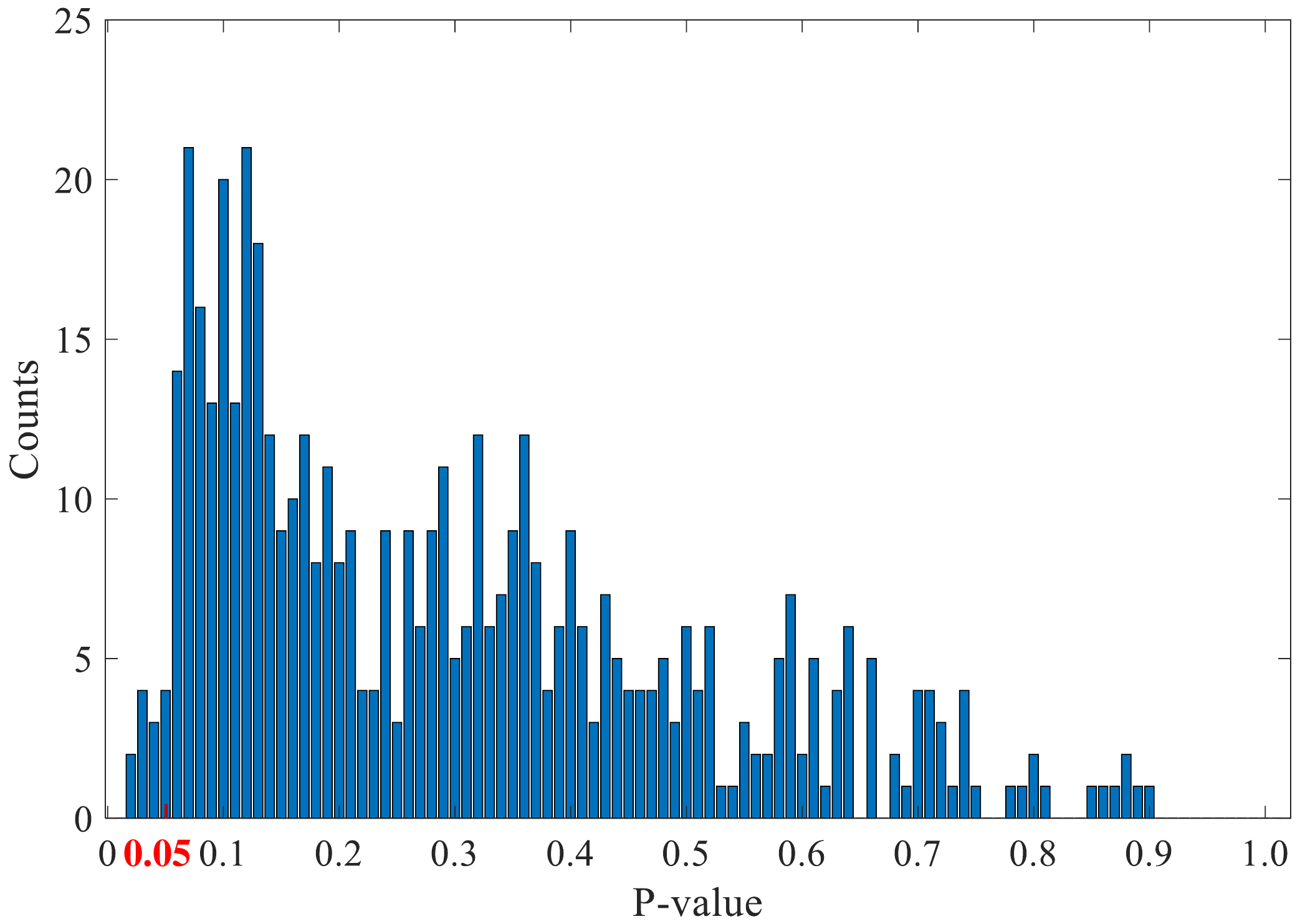}
		\caption{Histogram of the P-values for all 500 images.}
		\label{fig_pvalue}
	\end{figure}
	
	Let us denote the critical points given by all participants for the $n$-th original image as ${\mathbf{L}}_n=[L_n^1,L_n^2,\cdots,L_n^{60}]$. For Gaussian-like distribution, we remove the outlier data points from ${\mathbf{L}}_n$ according to the well-known 3-$\sigma$ criterion which assumes that a set of test data only contains random errors, calculate it to obtain standard deviation, and determine a range according to a certain probability of $99.7\%$ \cite{Pauta}. It is considered that the error exceeds this interval is not a random error. That is, a specific element $L_n^m$ will be excluded if it does not satisfy ${\mu}_{{\mathbf{L}}_n}-3{\sigma}_{{\mathbf{L}}_n} \le L_n^m \le {\mu}_{{\mathbf{L}}_n}+3{\sigma}_{{\mathbf{L}}_n}$ where ${\mu}_{{\mathbf{L}}_n}$ and ${\sigma}_{{\mathbf{L}}_n}$ denote the mean value and standard deviation of ${\mathbf{L}}_n$, respectively. The identified outliers have been marked in orange. After outlier removal, ${\mathbf{L}}_n$ will be updated to be ${\mathbf{L}}_{n}^{'}$. We further calculate the mean value ${\mu}_{{\mathbf{L}}_n^{'}}$ and set the final critical point for the $n$-th original image as follows:
	\begin{equation}
		L_n=\lceil {\mu}_{{\mathbf{L}}_n^{'}} \rceil,
	\end{equation}
	where $\lceil \cdot \rceil$ represents the round up operation. A detailed data analysis on vote distribution of different images is summarized in Table \ref{tab_dataanalysis}.
	
	\begin{figure}[!t]
		\centering
		\includegraphics[width=0.95\linewidth]{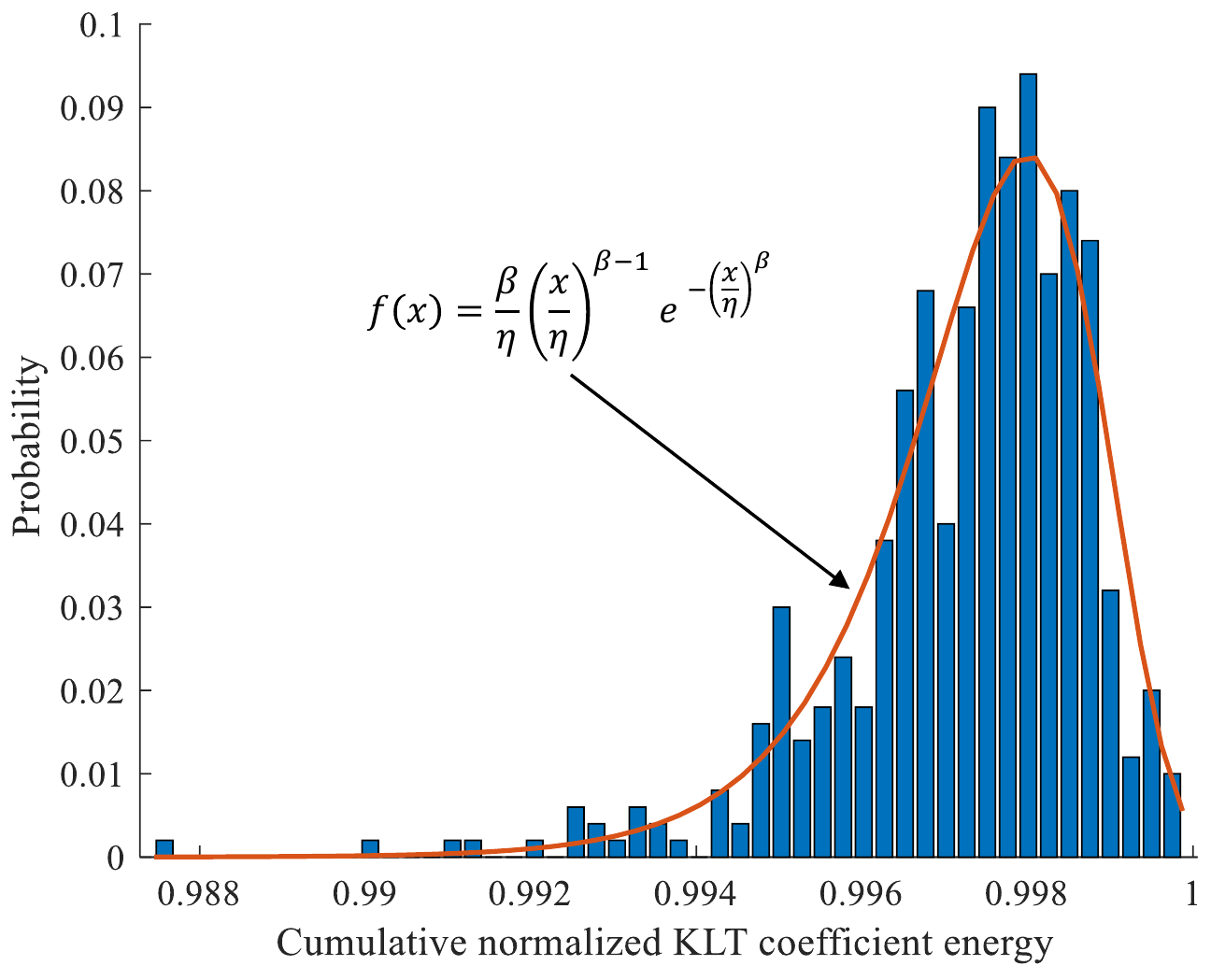}
		\caption{Visualization of the distribution of cumulative normalized KLT coefficient energy over all 500 original images. The fitted parameters are $\beta=894.16$ and $\eta=0.998$.}
		\label{fig_cumulativedistribution}
	\end{figure}
	
	In order to understand whether the vote distributions for all 500 images follow the normal distribution or not, we further conduct a normality test. Here, the Shapiro-Wilk test (i.e., SW test) \cite{SWtest} is adopted. The significance level is set to $\alpha$=0.05 \cite{SWtest}. If the P-value is less then $\alpha$, it means the sample data is significantly different from the normal distribution. Fig. \ref{fig_pvalue} shows the histgram of all 500 P-values. It is observed that only 13 out of all P-values were less than 0.05, only accounting for 2.6$\%$. This demonstrates that the vote distributions of most images (97.4$\%$) conform to normal distribution. Thus, our strategy using the 3-$\sigma$ criterion to filter the outliers is rational and appropriate.

	\begin{figure*}[!t]
		\centering
		\includegraphics[width=\textwidth]{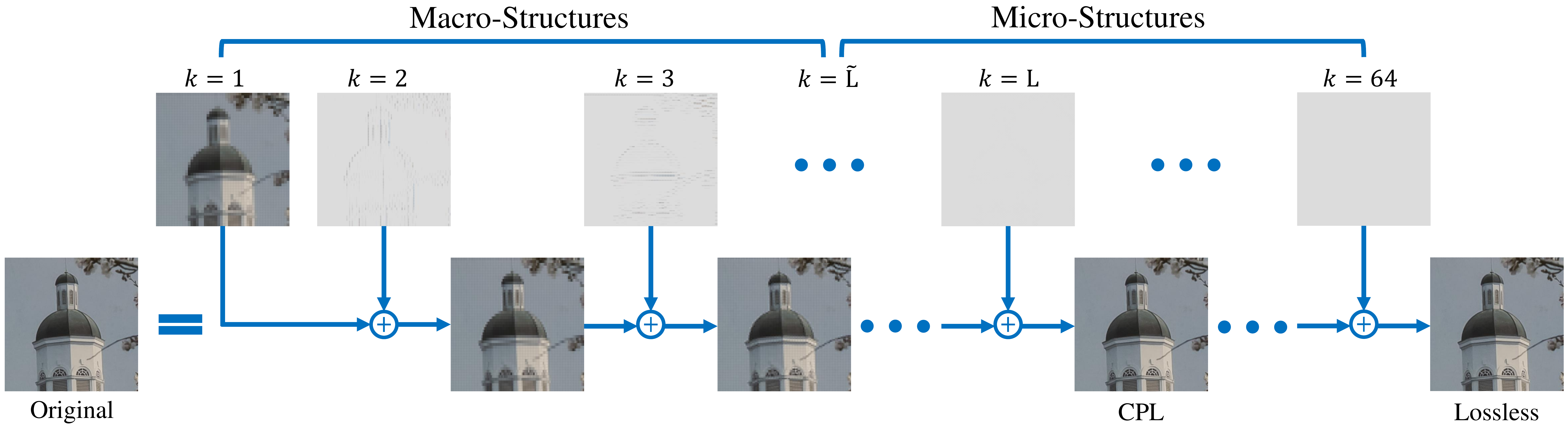}
		\caption{Visualization of the progressive image reconstruction process via inverse KLT.}
		\label{fig_reconstruction}
	\end{figure*}
	
	Next, we calculate the corresponding cumulative KLT coefficient energy $P_{L_n}$ for each original image with the obtained critical points $L_n\in\{1,2,\cdots,K\}$. As a result, we can obtain a vector ${\mathbf{P}}_L=[P_{L_1},P_{L_2},\cdots,P_{L_{500}}]$ with each element representing the critical cumulative KLT coefficient energy of a specific original image. The calculated cumulative KLT coefficient energy values of the same four images are given in the last column in Table \ref{tab_dataanalysis}. The distribution of cumulative normalized KLT coefficient energy over all 500 original images is also presented in Fig. \ref{fig_cumulativedistribution}. It is found that, although the critical points vary with different image contents, their corresponding cumulative KLT coefficient energy values tend to highly concentrate on the range of [0.99,1]. We can see from the figure that the distribution is right-skewed and can be well approximated by a Weibull distribution \cite{Weibull}. The Weibull distribution is a continuous probability distribution typically used for fitting both left- and right-skewed data. The widely used two-parameter Weibull distribution is expressed as follows:
	\begin{equation}
		f(x)=\frac{\beta}{\eta}\left(\frac{x}{\eta}\right)^{\beta-1} e^{-\left(\frac{x}{\eta}\right)^{\beta}}, \beta>0, \eta>0, x>0
	\end{equation}
	where $\beta$ and $\eta$ are the shape parameter and scale parameter, respectively. The above distribution can be well fitted by the Weibull distribution with $\beta=894.16$ and $\eta=0.998$. The fitted Weibull distribution curve (the red curve in Fig. \ref{fig_cumulativedistribution}) can be considered as a statistical prior which directly shows the probability of each $k$ (the index of spectral component) to be determined as the critical point.
	
	Given a test image, the expected value of $k$ is calculated to derive its corresponding critical point where the probabilities are determined by the fitted Weibull distribution function:
	\begin{equation}
		L=\bigg\lceil\frac{\sum_{k=1}^{K} k \cdot f\left(P_{k}\right)}{\sum_{k=1}^{K} f\left(P_{k}\right)}\bigg\rceil,
	\end{equation} 
	where $\lceil \cdot \rceil$ represents the round up operation.
	
	When we take the first $L$ spectral components for image reconstruction via inverse KLT, it is expected to generate the CPL counterpart of the original image. How the CPL image can be reconstructed with the first $L$ spectral components will be detailed in the following subsection. 
	
	\subsubsection{CPL Image Reconstruction}
	To utilize the first $L$ spectral components for image reconstruction via inverse KLT, we first define the corresponding reconstruction KLT coefficient matrix $\mathbf{Y}^{(L)}$ as follows:
	\begin{equation}
		\mathbf{Y}^{(L)} =
		\left[
		\begin{array}{cccccc}
			y_{1,1} & \cdots & y_{L,1} & 0 & \cdots & 0 \\
			\vdots & \ddots & \vdots & \vdots & \ddots & \vdots \\
			y_{1,S} & \cdots & y_{L,S} & 0 & \cdots & 0 \\
		\end{array}
		\right]^{\mathrm{T}}
	\end{equation}
	where $\mathbf{Y}^{(L)} \in \mathbb{R}^{K \times S},\;k=1,2,\cdots,K$. Then, the image is reconstructed as follows:
	\begin{equation}
		\mathbf{X}^{(L)} = \mathbf{P} \mathbf{Y}^{(L)}
	\end{equation}
	where $\mathbf{X}^{(L)}$ represents the reconstructed image by only considering the first $k$ spectral components. Since $L$ is the estimated critical point, $\mathbf{X}^{(L)}$ can be considered as the CPL image. Fig. \ref{fig_reconstruction} illustrates the progressive image reconstruction process. In our experiment, we set $K=64$. The images shown in the top row are the reconstructed results using only each individual single spectral component while the images shown in the bottom row are the reconstructed results using all previous spectral components. It is obvious that only using the first spectral component for reconstruction could recover most macro-structures of the original image. When more spectral components are involved, the image is progressively reconstructed with richer and finer details. Suppose $\tilde{L}$ is the boundary between macro- and micro-structures. Using the former $\tilde{L}$ spectral components will successfully reconstruct all the macro-structures and the latter $(64-\tilde{L})$ spectral components will mainly responsible for the reconstruction of micro-structures. For the critical point $L$ we have estimated, it means that we at least need to use the former $L$ spectral components to reconstruct all the macro-structures and sufficient micro-strucutres required for a perceptual lossless image. Finally, if all the spectral components are involved, the original image can be completely reconstructed.
	
\begin{figure}[!t]
		\centering
		\includegraphics[width=\linewidth]{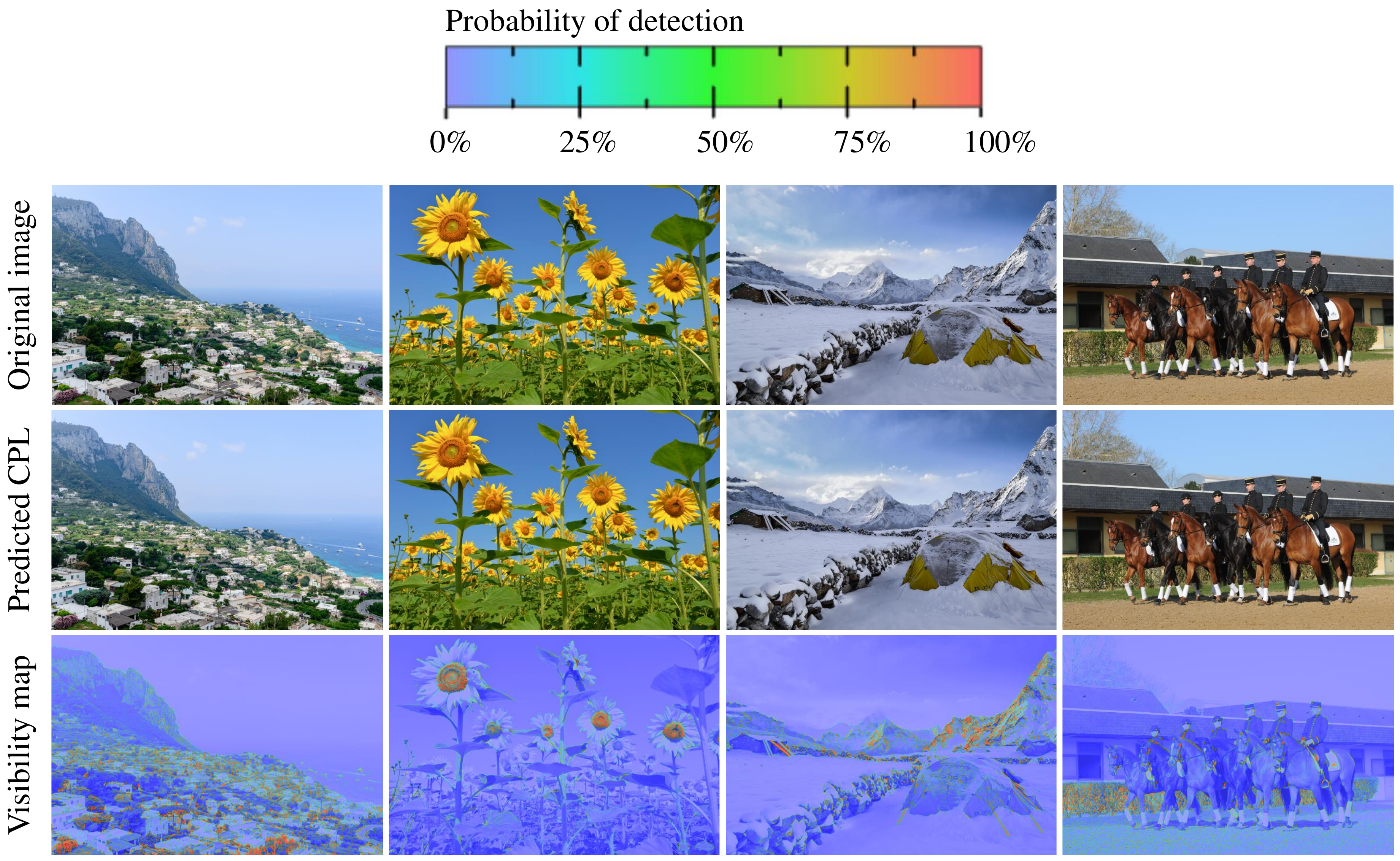}
		\caption{Visualization of the predicted CPL images and the  distortion visibility maps estimated by HDR-VDP2.2 \cite{HDR-VDP2.2}.}
		\label{fig_visibilitymap}
	\end{figure}

	In Fig. \ref{fig_visibilitymap}, we present our predicted CPL images of the four original images. As we can see, the CPL images (second row) are quite similar to the original ones, without obvious visible distortions. We also use the HDR-VDP2.2 metric \cite{HDR-VDP2.2} to calculate the distortion visibility maps, as shown in the third row. It is observed that most regions in the visibility maps are blue, implying that it is hard to perceive distortions from these CPL images. These results
	show that our determined CPL images are close to their original counterparts from the perspective of human perception.
	
	\subsubsection{JND Map Estimation}
	Once the CPL image ${\mathbf{X}}^{(L)}$ is obtained, we compute the difference map between the original image $\mathbf{X}$ and ${\mathbf{X}}^{(L)}$ as the final JND map $\mathbf{M}$:
	\begin{equation}
		{\mathbf{M}}(i,j) = \lvert {\mathbf{X}}(i,j) - {\mathbf{X}}^{(L)}(i,j) \rvert,
	\end{equation}
	where $\lvert \cdot \rvert$ denotes the absolute value operator, $(i,j)$ are the pixel coordinates in spatial domain.
	
	\section{Experiments}
	In this section, the proposed JND model is compared with the existing JND models to demonstrate its accuracy in direct prediction of JNDs and efficiency in eatimation of scene complexity and perceptual redundancy. 
	
	\subsection{Performance Comparison on Direct JND Prediction}
	In order to directly show the model ability to predict JND, we use the dataset in \cite{viewing-condition1} for validation. This dataset provides 20 reference images (i0webp-i19webp) along with their corresponding VLT data under different viewing conditions. According to the viewing conditions setting in our study, we select the VLT data measured under the viewing condition of 220$\rm cd/m^2$ peak luminance and 30 ppd. For each reference image in the dataset, we find the visually lossless version based on the provided VLT data. Thus, the difference map between the reference image and the visually lossless version is considered as the ground-truth JND map. Then, for each reference image, we also predict its JND maps by different JND models. Then, we conduct a normalization process on the predicted JND map and the ground-truth JND map, respectively. Specifically, for each pixel in the JND map, we divide it by its maximum value in that map and all pixel values will fall into the range of [0,1]. After normalization, we calculate the RMSE value between the predicted JND map and the ground-truth JND map. A smaller RMSE value means a higher similarity between the predicted JND map and the ground-truth JND map, thus directly showing a better ability of a specific model to predict JND. The experimental results are listed in Table \ref{tab_direct} where the best performer (smallest RMSE) for each image has been bolded. We can see that our proposed JND model owns the smallest RMSE values on all test images, demonstrating the best ability to predict JND.
	
	\begin{table*}[!t]
		\centering
		\footnotesize
		\caption{RMSE values between the predicted JND maps and the ground-truth JND maps. A smaller RMSE value means a better performance.}
		\label{tab_direct}
		\newcommand{\tabincell}[2]{\begin{tabular}{@{}#1@{}}#2\end{tabular}}
		\setlength{\tabcolsep}{4pt}
		\renewcommand\arraystretch{1.1}
		\begin{tabular}{c|c|c|c|c|c|c|c|c}
			\hline \hline
			\multirow{1}{*}{Image} & Yang2005 \cite{Yang2005} & Zhang2005 \cite{Zhang2005} & Wu2013 \cite{Wu2013} & Wu2017 \cite{Wu2017} 
			& Jakhetiya2018 \cite{JAKHETIYA} & Chen2020 \cite{Chen2020} & Shen2021 \cite{Shen2021} & Proposed  \\  \hline \hline
			i0webp & 0.3865 & 0.2408& 0.1521 &0.1918  &0.1822 &	0.2611&	0.2476 &\textbf{0.1032}   \\  \hline
			i1webp & 0.5284 & 0.2614& 0.1321 &0.1809  &0.2011 &	0.3749&	0.2530 &\textbf{0.0972}     \\  \hline
			i2webp & 0.3714 & 0.2625& 0.1580 &0.2849  &0.2099 &	0.2234&	0.2197 &\textbf{0.1447}   \\  \hline
			i3webp & 0.2255 & 0.1956& 0.1063 &0.1199  &0.1072 &	0.2522&	0.2024 &\textbf{0.0816}    \\  \hline
			i4webp & 0.5479 & 0.2567& 0.1580 &0.1992  &0.1730 &	0.3412&	0.2667 &\textbf{0.0795} \\  \hline
			i5webp & 0.6412 & 0.2813& 0.1741 &0.2422  &0.2334&	0.4649&	0.2201 & \textbf{0.0982}    \\  \hline
			i6webp & 0.3689&	0.2409&	0.1264 &0.1633 &0.1651&	0.2815 &0.2070 &\textbf{0.1202}  \\  \hline
			i7webp & 0.3151 &0.1995  & 0.1207 & 0.1852 & 0.1842 & 0.2350 & 0.2540 & \textbf{0.1017} \\ \hline
			i8webp & 0.4960 &0.2535 &0.1381 & 0.1774     &0.1923 &0.2742 &0.2853  & \textbf{0.1075}  \\  \hline 
			i9webp &0.3420 &0.3196 &0.1568 & 0.1665    &0.2021 &0.2728 &0.2114 & \textbf{0.1228}   \\  \hline
			i10jpeg &0.3290	&0.2383	&0.1778  &0.1841	&0.2516	&0.1923	&0.2066	&\textbf{0.1172}  \\  \hline
			i11jpeg & 0.3526 &0.1912	&0.0899 &0.1684	& 0.1151 & 0.2477	&0.2147 &\textbf{0.0743}	 \\  \hline
			i12jpeg &	0.3538&	0.2039&0.1636 &0.3025 &	0.2360&	0.2666&	0.1890 &\textbf{0.1051}  \\  \hline
			i13jpeg &	0.2423&0.2049&	0.0985 &0.1099 &	0.1020&	0.2010&	0.1440 &\textbf{0.0502}   \\  \hline
			i14jpeg &0.3078 &0.2822 &0.0960 &0.1293 &0.1273 &0.2371 &0.1555 &\textbf{0.0713} \\ \hline
			i15jpeg &0.2280	&0.2208	&0.0906	&0.1216 &0.1094	&0.2786	&0.2345	&\textbf{0.0378}  \\  \hline
			i16jpeg &0.3401	&0.2311	&0.1244 &0.1638	&0.1526	&0.2407	&0.2284	&\textbf{0.0949}  \\  \hline
			i17jpeg &0.3521 &0.2781	&0.1190 &0.1646	&0.1689 &0.3218	&0.1889 &\textbf{0.0790}	 \\  \hline
			i18jpeg &0.3990 &0.2285 &0.1049 &0.1541 &0.1706 &0.3074 &0.2524 &\textbf{0.0779}  \\  \hline
			i19jpeg &0.4718 &0.2697 &0.0901 &0.1268 &0.1693 &0.3359 &0.2400 &\textbf{0.0678}   \\  \hline
			Average &0.3800 &0.2430 &0.1289 &0.1768 &0.1727 &0.2805 &0.2211 &\textbf{0.0916}\\ \hline \hline
		\end{tabular}
	\end{table*}
	
	\begin{figure}[!t]
		\centering
		\includegraphics[width=\linewidth]{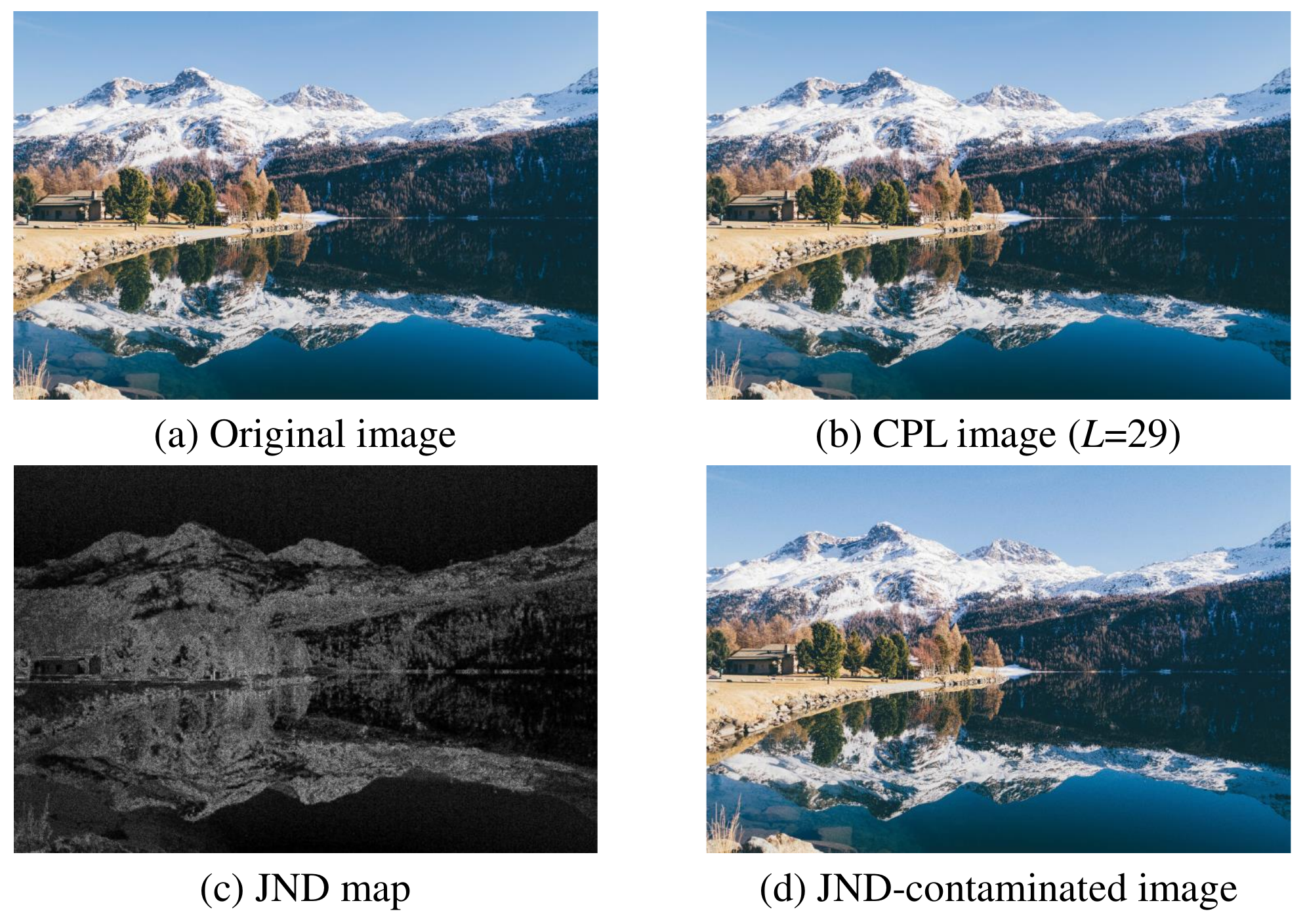}
		\caption{An example of noise contaminated image guided by our JND map. The PSNR is 26dB.}
		\label{fig_jndexample}
	\end{figure}
	
	\subsection{JND-Guided Noise Injection}
	According to previous works \cite{Wu2013,Wu2017,JAKHETIYA,Chen2020,Shen2021}, the performance of a JND model can be measured by the capability of hiding noise to some extent. Specifically, we can get the JND-contaminated image by injecting random noise into the image with the guidance of the estimated JND map as follows:
	\begin{equation}
		{\widetilde{\mathbf{X}}}(i,j) = {\mathbf{X}}(i,j)+\theta \cdot {\mathbf{N}}(i,j) \cdot {\mathbf{M}}(i,j),
	\end{equation}
	where ${\widetilde{\mathbf{X}}}$ is the JND-contaminated image, ${\mathbf{N}}$ denotes the bipolar random noise of $\pm1$, and $\theta$ is a noise energy regulating factor. By adjusting the value of $\theta$, the amount of noise injected into the image can be well controlled. In other words, we can inject almost the same amount of noise into the test image (almost the same PSNR) by adjusting the value of $\theta$ for different JND maps.
	
	In Fig. \ref{fig_jndexample}, we give an example to show the results of nosie contaminated image guided by our proposed JND map. Obviously, it is difficult to perceive the visual quality difference between the original image (i.e., Fig. \ref{fig_jndexample}(a)) and the contaminated image (i.e., Fig. \ref{fig_jndexample}(d)). That is, although the PSNR value is only 26dB, the injected noise are not visible. From the JND map shown in Fig. \ref{fig_jndexample}(c), we observe that many details in the complex textured regions are redundant to the HVS. Using such a JND map as a guidance for noise injection, the contaminated image will be perceived with high quality since most of the noises are encouraged to be added in those highly redundant area that is not sensitive to the HVS. 
	
	\begin{figure}[!t]
		\centering
		\includegraphics[width=\linewidth]{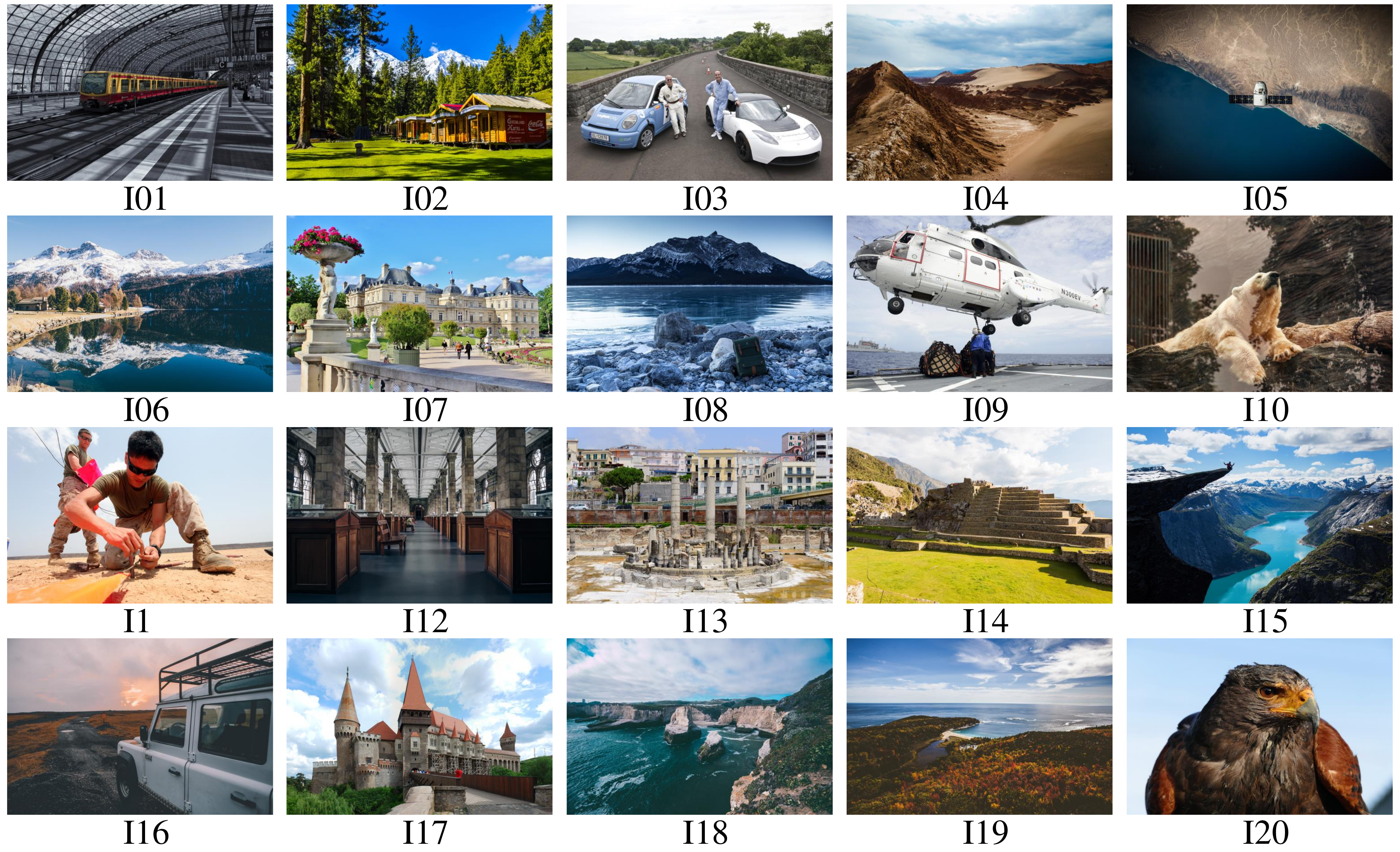}
		\caption{Twenty images used for testing in the experiments.}
		\label{fig_testimages}
	\end{figure}
	
	\begin{table*}[!t]
		\centering
		\scriptsize
		\caption{Performance comparison of different JND models on noise injection. The HDR-VDP score is based on HDR-VDP2.2 \cite{HDR-VDP2.2}.}
		\label{tab_noise}
		\setlength{\tabcolsep}{2pt}
		\renewcommand\arraystretch{1.5}
		\begin{tabular}{c|cc|cc|cc|cc|cc|cc|cc|cc}
			\hline \hline
			\multirow{2}{*}{Image}  & \multicolumn{2}{c|}{Yang2005 \cite{Yang2005}} & \multicolumn{2}{c|}{Zhang2005 \cite{Zhang2005}} & \multicolumn{2}{c|}{Wu2013 \cite{Wu2013}} & \multicolumn{2}{c|}{Wu2017 \cite{Wu2017}} & \multicolumn{2}{c|}{Jakhetiya2018 \cite{JAKHETIYA}} & \multicolumn{2}{c|}{Chen2020 \cite{Chen2020}} & \multicolumn{2}{c|}{Shen2021 \cite{Shen2021}} & \multicolumn{2}{c}{Proposed} \\ \cline{2-17}
			& HDR-VDP & MOS  & HDR-VDP  & MOS & HDR-VDP & MOS & HDR-VDP & MOS & HDR-VDP & MOS & HDR-VDP & MOS & HDR-VDP & MOS  & HDR-VDP& MOS\\ \hline \hline
			I01     & 55.8219 & -0.8333 & 54.4010 & -0.6333 & 56.6592 & -0.6000 & 58.3114 & -0.5667 & 58.7397 & -0.5778 & 55.8872 & -0.3889 & \textbf{58.9145} &-0.6333  & 58.1479 & \textbf{-0.2313}  \\ \hline
			I02     & 56.0151 & -0.6444 & 54.4634 & -0.7222 & 58.0067 & -0.4667 & 59.3012 & -0.3889 & 59.5552 & -0.3778 & 53.7419 & -0.6222 & 54.5701 &-0.5778   & \textbf{59.8919} & \textbf{-0.1938}  \\ \hline
			I03     & 54.2115 & -0.6556 & 53.4778 & -0.7333 & 56.1152 & -0.6000 & 57.8486 & -0.5333 & 56.4067 & -0.5889 & 54.3552 & -0.6778 & 53.7209 &-0.5889  & \textbf{58.2178} & \textbf{-0.2125}  \\ \hline
			I04     & 53.7742 & -0.5556 & 52.1998 & -0.7111 & 55.0556 & -0.5111 & 57.0854 & -0.3556 & 55.7968 & -0.4111 & 52.5509 & -0.8222 & \textbf{58.3130} &-0.3778  & 56.9190 & \textbf{-0.1938}  \\ \hline
			I05     & 52.3229 & -0.7111 & 50.4194 & -0.6222 & 52.1538 & -0.5778 & 54.2264 & -0.4222 & 52.9179 & -0.4444 & 52.1209 & -0.5889 & \textbf{57.3850} &-0.3778 & 57.2651 & \textbf{-0.2563}  \\ \hline
			I06     & 52.3135 & -0.6222 & 52.5651 & -0.5444 & 54.9231 & -0.5000 & 56.2502 & -0.3556 & 56.4490 & -0.3889 & 52.3666 & -0.7222 & \textbf{58.6321} &-0.3444 & 57.0403 & \textbf{-0.2438}  \\ \hline
			I07     & 56.2070 & -0.6333 & 54.7842 & -0.6333 & 56.3356 & -0.5222 & \textbf{60.3065} & -0.4111 & 58.4491 & -0.3556 & 52.5698 & -0.7111 & 54.0611 &-0.6333  & 58.7755 & \textbf{-0.2563}  \\ \hline
			I08     & 53.9325 & -0.4778 & 52.7390 & -0.5778 & 55.3771 & -0.5222 & 57.7465 & -0.2889 & 56.5790 & -0.5667 & 52.6583 & -0.7222 & 55.0882 &-0.3778  & \textbf{58.6738} & \textbf{-0.2375}  \\ \hline
			I09     & 54.2341 & -0.5667 & 52.5777 & -0.6444 & 54.6901 & -0.5333 & 56.8661 & -0.3778 & 54.1190 & -0.4889 & 52.5546 & -0.7444 & 52.5290 &-0.6333 & \textbf{56.9770} & \textbf{-0.2688}  \\ \hline
			I10     & 55.1364 & -0.6000 & 53.0932 & -0.6667 & 55.3694 & -0.5222 & \textbf{57.1095} & -0.4111 & 56.4227 & -0.3444 & 54.8467 & -0.5333 & 54.4159 &-0.3556  & 56.5825 & \textbf{-0.2125}  \\ \hline
			I11     & 53.0243 & -0.6000 & 51.6517 & -0.6000 & 54.2494 & -0.4778 & 56.0900 & -0.3000 & 53.6028 & -0.4000 & 51.9667 & -0.7222 & 52.5632 &-0.6333  & \textbf{57.0014} & \textbf{-0.2438}  \\ \hline
			I12     & 53.4857 & -0.4333 & 51.0445 & -0.4000 & 53.4636 & -0.3444 & 54.4890 & -0.3778 & 54.4000 & -0.3000 & 53.1855 & \textbf{-0.2778} & 51.6240 &-0.3556  & \textbf{55.5404} & -0.3438  \\ \hline
			I13     & 58.0653 & -0.5556 & 56.2722 & -0.5778 & 58.6279 & -0.4778 & \textbf{61.3581} & -0.3667 & 60.1869 & -0.4000 & 52.1352 & -0.6778 & 55.7032 &-0.6778  & 60.3400 & \textbf{-0.2688}  \\ \hline
			I14     & 55.5298 & -0.3667 & 54.5143 & -0.4667 & 57.3347 & -0.4111 & 59.8811 & -0.3000 & 56.2978 & -0.4889 & 51.9380 & -0.7889 & 57.5791 &-0.6444  & \textbf{60.5972} & \textbf{-0.2375}  \\ \hline
			I15     & 56.0810 & -0.4111 & 53.1295 & -0.4889 & 56.3243 & -0.4222 & \textbf{57.4761} & -0.3000 & 56.4456 & -0.4556 & 54.0723 & -0.7444 & 56.7220 &-0.5778  & 57.3237 & \textbf{-0.2125}  \\ \hline
			I16     & 50.9370 & -0.3333 & 50.3087 & -0.5333 & 51.8376 & -0.5000 & 53.2925 & -0.3444 & 52.6953 & -0.3778 & 51.2403 & -0.6333 & 51.1785 &-0.7333  & \textbf{54.9153} & \textbf{-0.2875}  \\ \hline
			I17     & 53.1796 & -0.4222 & 53.4795 & -0.6222 & 55.1712 & -0.4667 & \textbf{57.0430} & -0.3111 & 55.8383 & -0.4000 & 54.0341 & -0.6778 & 52.4972 &-0.3556   & 56.6582 & \textbf{-0.2750}  \\ \hline
			I18     & 52.2712 & -0.4222 & 52.4559 & -0.5444 & 54.4790 & -0.3778 & 58.0745 & -0.2333 & 56.3035 & -0.3000 & 52.0321 & -0.6333 & 51.3789 &-0.5778  & \textbf{59.1816} & \textbf{-0.2625}  \\ \hline
			I19     & 52.1117 & -0.3778 & 52.1802 & -0.5889 & 54.3710 & -0.4333 & 57.6392 & -0.3222 & 55.7807 & -0.3778 & 52.1641 & -0.7111 & 57.1253 &-0.5000  & \textbf{58.1924} & \textbf{-0.2375}  \\ \hline
			I20     & 53.6226 & -0.3778 & 51.4270 & -0.5444 & 54.2392 & -0.4667 & 56.1542 & -0.2889 & 54.4725 & -0.3667 & 52.1627 & -0.6333 & 49.4429 &-0.6333  & \textbf{56.7589} & \textbf{-0.2438}  \\ \hline
			Average & 54.1139 & -0.5300 & 52.8592 & -0.5928 & 55.2392 & -0.4867 & 57.3275 & -0.3628 & 56.0729 & -0.4206 & 52.9292 & -0.6517 & 54.6722 &-0.5294  & \textbf{57.7500} & \textbf{-0.2459}  \\ \hline \hline
		\end{tabular}
	\end{table*}
	
	\subsubsection{Performance Comparison on JND-Guided Noise Injection} Apparently, with the guidance of a more accurate JND map, the JND-contaminated image (obtained by Eq. (13)) with same noise level should have better visual quality. As stated, for different JND models, we can adjust the value of $\theta$ to ensure that almost the same amount of noise is injected. Then, the performances of differenta JND model can be compared objectively and subjectively. Specifically, objective evaluation is performed based on the HDR-VDP2.2 metric \cite{HDR-VDP2.2} which is known as a popular distortion visibility detection metric for both high-dynamic range and standard images. A higher HDR-VDP score indicates less visible distortions. Subjective evaluation is performed by humans to obtain MOS score. Another 30 participants are invited to participate the subjective experiments. Each participant is asked to assign an opinion score in the range [-1,0] to a specific image pair including one original image and one JND-contaminated image. A score equals to 0 means the visual quality of the JND-contaminated image is equal to that of the original image while a score equal to -1 means the visual quality of the JND contaminated image is dramatically worse than that of the original image. As a result, for each contaminated image, 30 scores are collected from all participants. After removing the outlier, the mean value of the retained opinion scores is calculated as the MOS. Overall, a more accurate JND model will result in higher HDR-VDP score and MOS than the competitors.

	We select another 20 high-quality images from the DIV2K \cite{DIV2K1,DIV2K2} dataset for testing. These images are shown in Fig. \ref{fig_testimages}. The proposed JND model is compared with seven existing JND models including Yang2005 \cite{Yang2005}, Zhang2005 \cite{Zhang2005}, Wu2013 \cite{Wu2013}, Wu2017 \cite{Wu2017}, Jakhetiya2018 \cite{JAKHETIYA}, Chen2020 \cite{Chen2020}, and Shen2021 \cite{Shen2021}. Table \ref{tab_noise} provides the performance results of different JND models in terms of HDR-VDP score and MOS at the noise level of PSNR=26dB (comparisons under other different PSNR settings are also conducted and will be reported later in this section). It can be seen that our proposed JND model delivers higher HDR-VDP scores on 11 out of 20 test images and the highest HDR-VDP score in average when considering all 20 images. Followed by our proposed JND model, Wu2017 and Shen2021 take the first place for 5 and 4 times, respectively. In terms of MOS, we achieve the highest MOS values for almost all the images except the image `I12' on which Chen2020 is the best. These results demonstrate the superiority of our proposed JND model against others in allocating noise at the noise level of PSNR=26dB (comparisons under other different PSNR settings are also conducted and will be reported later in this section). Note that we inject noise to the original images at the noise level of PSNR=26dB to obtain the JND-guided noise contaminated images. Generally, 26dB is a relatively large noise intensity which will probably break the transparency, and thus it is inevitable to induce visible distortions on the JND-guided noise contaminated image to some degree.
	
	\begin{figure}[!t]
		\centering
		\includegraphics[width=\linewidth]{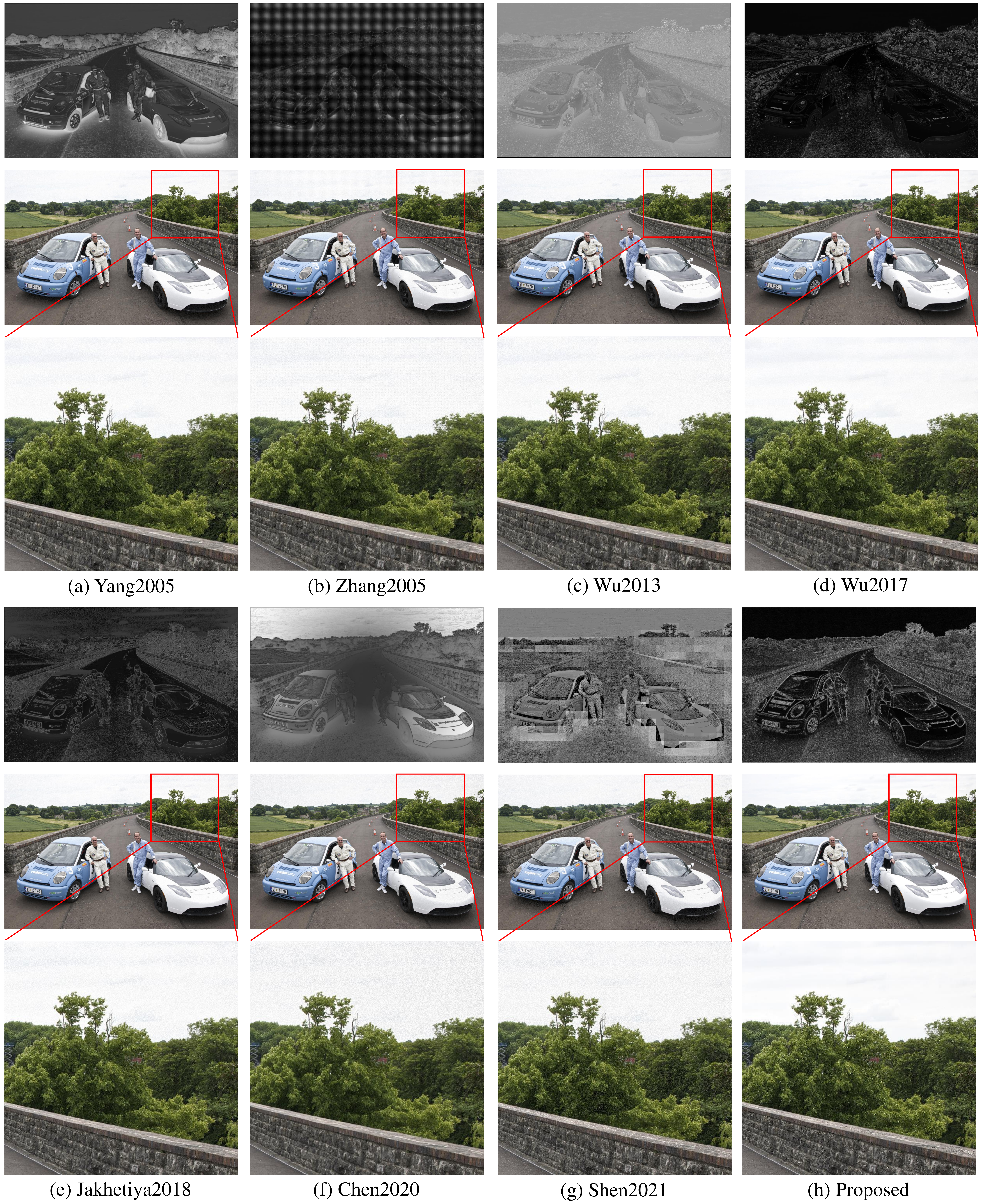}
		\caption{Visual comparison of noise-contaminated images generated by using different JND models as guidance. Zoom-in for best viewing.}
		\label{fig_results1}
	\end{figure}

	\begin{table*}[!t]
		\centering
		\footnotesize
		\caption{Performance comparison on maximum tolerable noise level measured by PSNR. A smaller PSNR value means a better performance.}
		\label{tab_noisetolerance}
		\newcommand{\tabincell}[2]{\begin{tabular}{@{}#1@{}}#2\end{tabular}}
		\setlength{\tabcolsep}{5pt}
		\renewcommand\arraystretch{1.2}
		\begin{tabular}{c|c|c|c|c|c|c|c|c}
			\hline \hline
			\multirow{1}{*}{Image} & Yang2005 \cite{Yang2005} & Zhang2005 \cite{Zhang2005} & Wu2013 \cite{Wu2013} & Wu2017 \cite{Wu2017} 
			& Jakhetiya2018 \cite{JAKHETIYA} & Chen2020 \cite{Chen2020} & Shen2021 \cite{Shen2021} & Proposed  \\  \hline \hline
			I01 & 36.3267 & 37.0200 &36.7256  &36.5450  &36.7110  &	37.1360 &38.6131&\textbf{35.5836}   \\  \hline
			I02 & 31.7161  & 30.2181 & 30.4085  &29.9859   &\textbf{28.8173} &	36.7045 &	31.4304  &30.9822      \\  \hline
			I03 & 34.5942  & 37.6568 & 34.7696 &36.8901   &36.7906  &	40.9209 & \textbf{33.4430} &34.4538   \\  \hline
			I04 & 32.0325  & 34.4976 & 31.7433  &31.0419   &33.2436  &	38.6290 &	36.8489  &\textbf{30.9284 }    \\  \hline
			I05 & 36.1455  & 34.9346 & 34.9324  &34.5491   &35.5152  &	39.4872 &	36.2114  &\textbf{34.3448 } \\  \hline
			I06 & 34.1929  & 34.5028 & 33.0136  &32.5747   &33.8537 &	36.0527 &	36.8339  & \textbf{32.3934 }    \\  \hline
			I07 & 32.2634 &	32.0573 &	\textbf{31.0054}  &31.7650  &31.3990 &	35.1954  &32.7384  &31.4912   \\  \hline
			I08 & 33.9132  &34.0813   & 31.9071  & 31.3348  & 35.5463  & 36.7112  & 43.1402  & \textbf{31.1622 } \\ \hline
			I09 & 37.8955  &33.9826  &  \textbf{32.5214}  & 38.6528      &37.9368  &38.2275  &42.7123   & 36.4163  \\  \hline 
			I10 &32.4285  &32.0577  &32.4657  & 31.8867     &31.9063  &33.4762  &33.7650  & \textbf{30.7900 }   \\  \hline
			I11 &38.1032 	&37.5925 	&32.7340   &\textbf{32.6700}	&34.9148 	&37.2067 &39.7867 	&35.8269   \\  \hline
			I12 & 35.7780  &37.4056 	&36.5880 &\textbf{35.4858 }	& 35.9974  & 38.0843 	&37.6373  &36.3014 	 \\  \hline
			I13 &	32.6346 &	32.6721 &32.8009  &32.5361  &	32.1162 &36.2254 &	\textbf{31.2969} &31.8158   \\  \hline
			I14 &31.8024 &34.7035 &	32.1759  &31.5463  &34.2748 &37.6211 &37.7132  &\textbf{30.8307}   \\  \hline
			I15 &32.3944  &34.3741  &\textbf{32.0731}  &33.0717  &33.4020  &34.7803  &40.2603  &33.4000  \\ \hline
			I16 &36.9723 	&37.1364 &\textbf{35.7520 } &36.0867  &37.7459 	&37.9674 	&41.6559 &36.3747   \\  \hline
			I17 &32.0907 	&40.6499 &31.0099  &31.0817 	&36.3353 	&34.4205 	&41.9413 	&\textbf{30.4702}  \\  \hline
			I18 &31.4816  &33.1370 	&33.0206  &30.8787 	&36.5084 &35.1940 	&35.7879  &\textbf{30.4573}	 \\  \hline
			I19 &32.0469  &30.8358  &31.5857  &30.8322  &32.3963  &34.5194  &35.4254  &\textbf{30.2432}  \\  \hline
			I20 &32.9795  &34.1478  &33.1078  &32.7811  &33.9243  &33.9742  &41.6431  &\textbf{32.3365 }   \\  \hline
			Average &33.8896  &34.6832  &33.0170  &33.1098  &34.4668  &36.6267  &37.4442  &\textbf{32.8301 }\\ \hline \hline
		\end{tabular}
	\end{table*} 
	
	In order to make a more clear comparison among these JND models, Fig. \ref{fig_results1} gives a visual example of the JND maps and the noise contaminated images. Images in the first row are the estimated JND maps. Images in the second row are the JND-guided noise-injected images generated according to Eq. (12). Images in the third row are the enlarged versions of sub-regions cropped from the noise-injected images in the second row. We can observe that the noise-contaminated image guided by our proposed JND model is the `cleanest' one among all the compared ones. It seems that Fig. \ref{fig_results1}(h) is perceptually the same with the original image as it seems to have no visible noise. Among the competitors, Fig. \ref{fig_results1}(d) also has achieved satisfactory visual quality while it is still worse than Fig. \ref{fig_results1}(h). With careful observations, one can still perceive visible noise in the sky area of Fig. \ref{fig_results1}(d). 
	
	By taking a closer look at Fig. \ref{fig_results1}(h), we find that the injected noises are hidden in those textured areas such as the walls and leaves. However, it is hard to be perceived by the HVS at the first glance. As has been reported in \cite{Wu2017}, the HVS is highly adapted to extracting the repeated patterns for visual content representation and it is hard to perceive the noise which is injected into the textured areas with high pattern complexity. By contrast, it is much easier for the HVS to perceive the noise in those relatively smooth areas. Compared with the smooth areas, textured areas tend to contain much more details (e.g., micro-structures) and are likely to be redundant. As we have demonstrated before, the difference map between the original image and the derived CPL image can well resemble the redundant micro-structure image information to the HVS. Thus, our proposed JND model using such difference map as the JND map will encourage to hide the noise in those highly redundant areas. Therefore, the injected noises will not be easily visible by the HVS. What's more, instead of modeling and aggregating the masking effects of diverse contributing factors in isolation, our proposed JND model dedicates to deriving a CPL image to best exploit the potential perceptual redundancies exist in the original image from a top-down perspective. Thus, our proposed JND model would be able to implicitly characterize the influence of more potential factors beyond those have been considered previously. As a result, our proposed JND model can finally generate a perceptually better noise-contaminated image. 
	
	\begin{figure}[!t]
		\centering
		\includegraphics[width=0.8\linewidth]{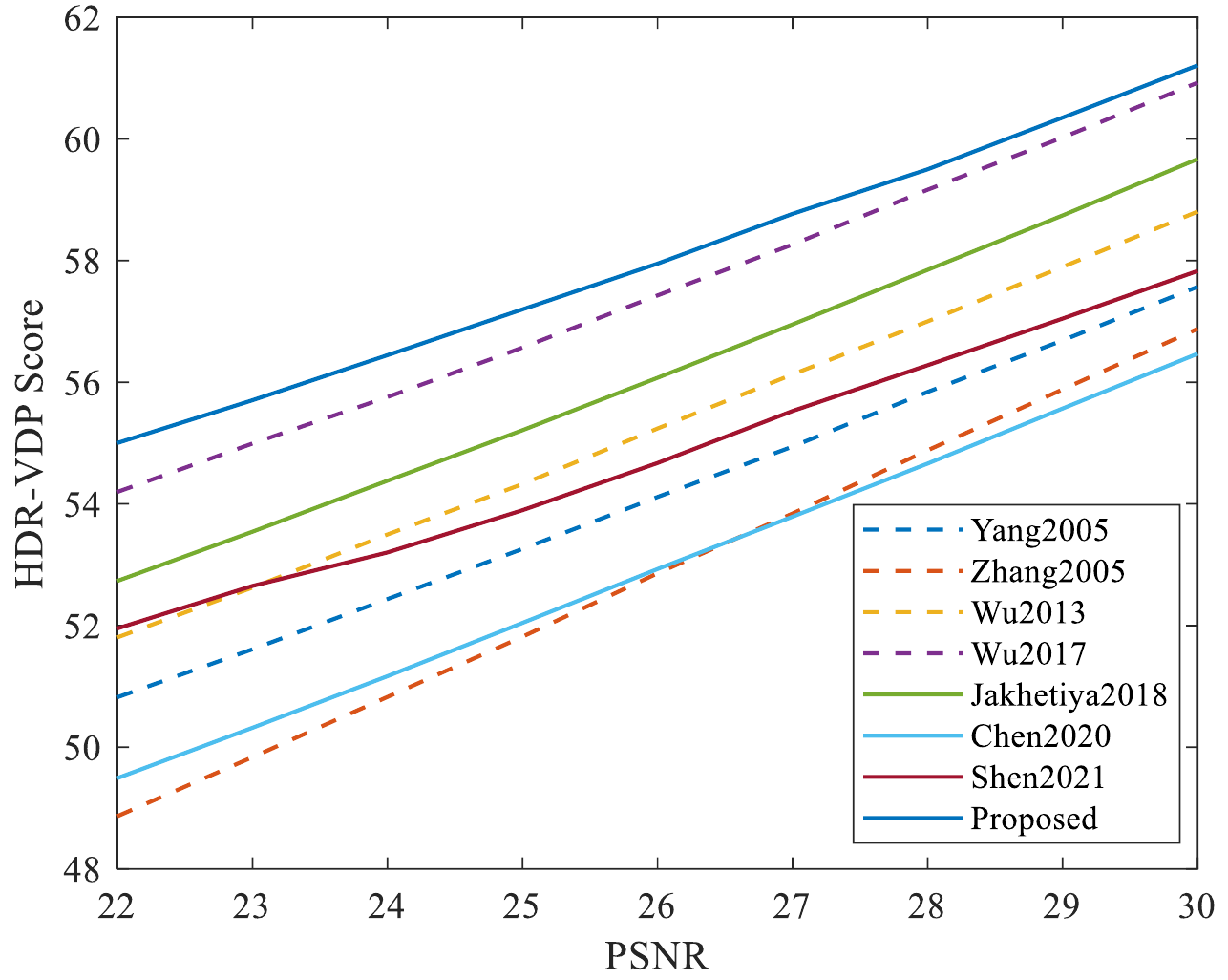}
		\caption{Curve of average HDR-VDP score versus PSNR.}
		\label{fig_morePSNR}
	\end{figure}
	
	Note that comparing the performance under PSNR=26dB only may be unfair to certain JND models. Therefore, we also conduct experiments under more PSNR settings. Specifically, we set PSNR=$\{22, 23, 24, 25, 26, 27, 28, 29, 30\}$ and for each PSNR setting we generate JND-contaminated images of I01-I20 by using different JND models as guidance. We also apply the HDR-VDP metric to compute the objective score of each JND-contaminated image, and we take the averaged HDR-VDP score as the indicator of JND model performance. Finally, we plot the curve of average HDR-VDP score versus PSNR value, as shown in Fig. \ref{fig_morePSNR}. We can see that the curve of our proposed JND model is among the top, which indicates our JND model always achieves the highest HDR-VDP score at each PSNR value.
	
	It should be emphasized that, although we have compared different JND models on noise injection with different noise levels, such experiments do not directly reflect the capability of predicting JND. The experimental setups for directly comparing the capability of predicting JND still deserve more rigorous treatments.
	
	\begin{figure*}[!t]
		\centering
		\includegraphics[width=\textwidth]{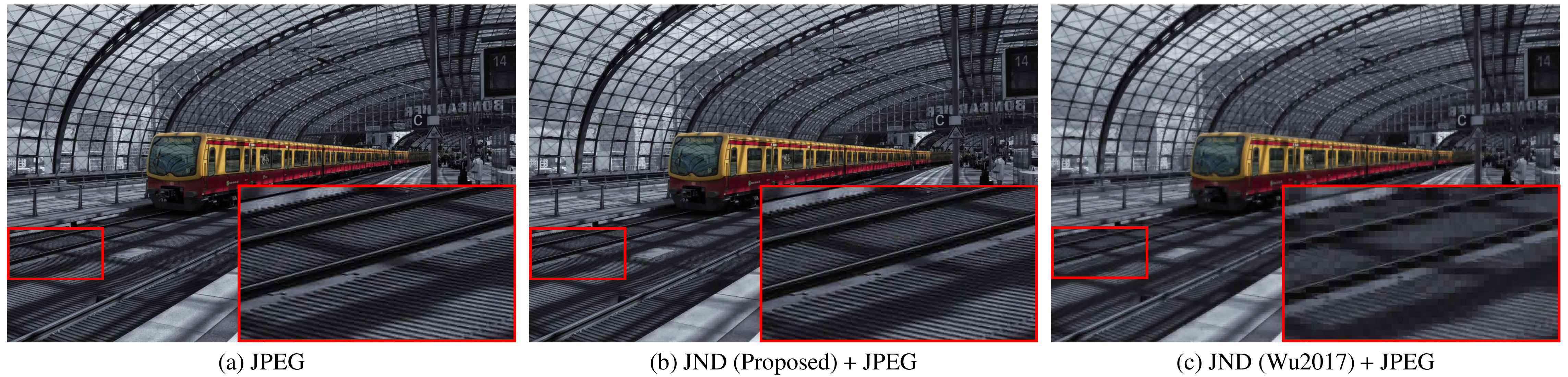}
		\caption{Visual quality comparison of JND-guided image compression. (a) Direct JPEG compression result; (b) Our proposed JND model-guided JPEG compression result; (c) Wu2017 JND model-guided JPEG compression result.}
		\label{fig_compression}
	\end{figure*}

	\subsubsection{Performance Comparison on Maximum Tolerable Noise Level} The above experiments are conducted to compare the visual quality of different JND-guided noise contaminated images under the same PSNR level. While the results have demonstrated promising performance of our proposed JND model, it is still necessary to understand the maximum tolerable noise level by different JND models. A better JND model should tolerate more noise while keeping the quality of the noise-contaminated image perceptually unchanged. In our experiment, we use PSNR as a measure of the noise level and a lower PSNR value actually indicates a better performance of a JND model. The results are listed in Table \ref{tab_noisetolerance}. As we can see, the proposed JND model can tolerable more noise on 11 out of 20 images and also has the lowest PSNR in average when considering all 20 images, which again validates the superiority of our proposed JND model.
	
	\begin{table}[!t]
		\centering
		\footnotesize
		\caption{Performance comparison on JND-guided JPEG compression (QF=1). A higher value of $G$ means a better performance.}
		\label{tab_compression}
		\setlength{\tabcolsep}{2.5pt}
		\renewcommand\arraystretch{1.3}
		\begin{tabular}{c|c|c|c|c|c|c}
			\hline \hline
			\multirow{2}{*}{Image} & \multicolumn{3}{c|}{JND (Proposed) + JPEG} & \multicolumn{3}{c}{JND (Wu2017) + JPEG} \\ \cline{2-7} 
			& $\Delta{Bitrate}$       & $\Delta{PSNR}$         & $G$       & $\Delta{Bitrate}$      & $\Delta{PSNR}$         & $G$     \\ \hline \hline
			I01                    & 9.07\%       & 2.44\%       & 3.7141      & 47.12\%      & 17.46\%      & 2.6988     \\ \hline
			I02                    & 4.83\%        & 1.08\%       & 4.4883      & 39.64\%      & 29.68\%      & 1.3357     \\ \hline
			I03                    & 23.24\%       & 4.86\%       & 4.7857      & 42.74\%      & 11.30\%      & 3.7819     \\ \hline
			I04                    & 15.65\%       & 3.55\%       & 4.4047      & 43.47\%      & 13.89\%      & 3.1287     \\ \hline
			I05                    & 23.83\%       & 4.99\%       & 4.7740      & 43.89\%      & 12.51\%      & 3.5099     \\ \hline
			I06                    & 10.05\%       & 2.19\%       & 4.5872      & 38.10\%      & 12.62\%      & 3.0183     \\ \hline
			I07                    & 11.44\%       & 3.19\%       & 3.5839      & 38.60\%      & 15.07\%      & 2.5610     \\ \hline
			I08                    & 13.62\%       & 4.19\%       & 3.2484      & 44.82\%      & 17.93\%      & 2.4996     \\ \hline
			I09                    & 13.43\%       & 4.93\%       & 2.7213      & 35.16\%      & 15.98\%      & 2.2006     \\ \hline
			I10                    & 24.45\%       & 5.30\%       & 4.6149      & 46.94\%      & 14.29\%      & 3.2859     \\ \hline
			I11                    & 14.66\%       & 5.05\%      & 2.9033     & 37.34\%      & 16.35\%      & 2.2839     \\ \hline
			I12                    & 10.25\%       & 3.01\%       & 3.3996      & 44.10\%      & 20.35\%      & 2.1671     \\ \hline
			I13                    & 13.33\%       & 3.58\%       & 3.7282      & 43.88\%      & 14.37\%      & 3.0536     \\ \hline
			I14                    & 25.92\%       & 4.61\%       & 5.6230      & 51.00\%      & 11.93\%      & 4.2754     \\ \hline
			I15                    & 18.70\%       & 3.24\%       & 5.7735      & 45.15\%      & 13.19\%      & 3.4214     \\ \hline
			I16                    & 15.82\%       & 4.50\%       & 3.5121      & 36.45\%      & 15.75\%      & 2.3143     \\ \hline
			I17                    & 24.62\%       & 5.14\%       & 4.7868      & 46.18\%      & 12.35\%      & 3.7384     \\ \hline
			I18                    & 16.80\%       & 3.32\%       & 5.0552      & 45.34\%      & 11.78\%      & 3.8485     \\ \hline
			I19                    & 21.54\%       & 3.25\%       & 6.6243      & 47.82\%      & 8.98\%       & 5.3267     \\ \hline
			I20                    & 19.65\%       & 4.89\%       & 4.0204      & 39.94\%      & 12.24\%      & 3.2626     \\ \hline
			Average                & 16.54\%       & 3.87\%       & 4.2793      & 42.88\%      & 14.90\%      & 2.8779     \\ \hline \hline
		\end{tabular}
	\end{table}
	
	\subsection{JND-Guided Image Compression}
	Since the JND map implies the visibility limitation of the HVS. Thus, it is often employed in image compression to improve compression efficiency. Generally, the smoothing operation can reduce the signal variance, which makes image compression easier. However, blindly smoothing operations will always jeopardize image quality. Thus, we can use the JND map to guide the smoothing operation as a preprocess step before JPEG compression for perceptual redundancy reduction. As will be illustrated later, the visual quality of the compressed image guided by our proposed JND model will not be affected too much while saving considerable coding bits. Specifically, given an input image $\mathbf{T}$ and its corresponding JND map ${\mathbf{T}}_{\text{M}}$, the JND-guided image smoothing operation is described as follows:
	\begin{equation}
		\footnotesize
		\widetilde{\mathbf{T}}(i,j)=\left\{
		\begin{array}{cc}
			\mathbf{T}(i,j)+\mathbf{T}_{\text{M}}(i,j), & \mathbf{T}(i,j)-\overline{\mathbf{T}}_{\text{P}}<-\mathbf{T}_{\text{M}}(i,j) \\	
			\mathbf{T}(i,j)-\mathbf{T}_{\text{M}}(i,j), & \mathbf{T}(i,j)-\overline{\mathbf{T}}_{\text{P}}>\mathbf{T}_{\text{M}}(i,j) \\
			\overline{\mathbf{T}}_{\text{P}}, & else
		\end{array} \right.
	\end{equation}
	where $\overline{\mathbf{T}}_{\text{P}}$ denotes the mean value of the block that $\mathbf{T}(i,j)$ belongs to during the compression process (e.g., the divided $8\times8$ blocks that $\mathbf{T}(i,j)$ located at during JPEG compression).
	
	With the above JND-guided image smoothing operation, the visual redundancy of the input image will be reduced to facilitate compression. One visual example is shown in Fig. \ref{fig_compression} where we perform JPEG compression of the original image ``I01'' at QF=1 in two different ways: direct JPEG compression and JND-guided JPEG compression. Fig. \ref{fig_compression}(a) is the result obtained by direct JPEG compression. Fig. \ref{fig_compression}(b) is the result obtained by first preprocessing the original image with the JND-guided image smoothing operation and then performing JPEG compression. It can be observed that, though less bit rate is required (0.0684 bpp for Fig. \ref{fig_compression}(a) and 0.0622 bpp for Fig. \ref{fig_compression}(b)), the visual quality of Fig. \ref{fig_compression}(b) is almost equal to that of Fig. \ref{fig_compression}(a). As a comparison, we also provide the result obtained by using Wu2017 JND model as the guidance under the same QF setting in JPEG compression. As shown in Fig. \ref{fig_compression}(c), though the bit rate of Fig. \ref{fig_compression}(c) is 0.0362 bpp which is less than Fig. \ref{fig_compression}(c), we can easily perceive obvious compression artifact in this result.
	
	\begin{figure}[!t]
		\centering
		\includegraphics[width=0.8\linewidth]{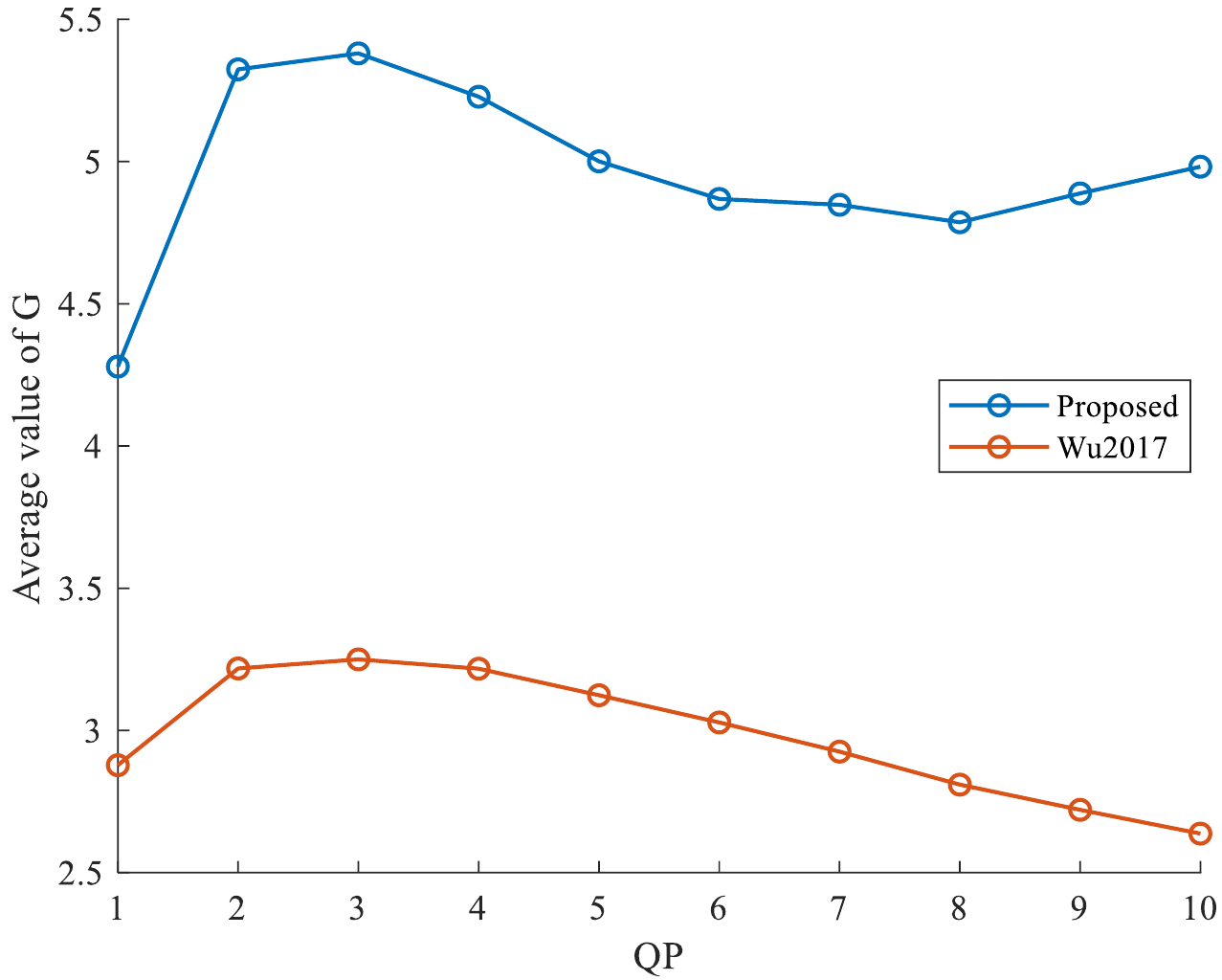}
		\caption{The averge value of $G$ with different QF settings in JPEG compression.}
		\label{fig_gain}
	\end{figure}
	
	We also notice that in comparison with Fig. \ref{fig_compression}(c), Fig. \ref{fig_compression}(b) achieves higher visual quality at the cost of higher bit rate. Therefore, an important issue is how to fairly compare the overall performance of different JND-guided JPEG compression results by jointly taking visual quality and bit rate into account with a single criteria. Keeping this in mind, we define the gain $\small G=\frac{\Delta{Bitrate}}{\Delta{PSNR}}$ as the criteria, where $\Delta{Bitrate}$ denotes the saving of bit rate and $\Delta{PSNR}$ denotes the reduction of PSNR value. Specifically, these two measures are calculated as follows:
	\begin{equation}
		\small
		\Delta{Bitrate}=\frac{{Bitrate}_{ori}-{Bitrate}_{jnd}}{{Bitrate}_{ori}}\times100\%
	\end{equation}
	and
	\begin{equation} 
		\small
		\Delta{PSNR}=\frac{{PSNR}_{ori}-{PSNR}_{jnd}}{{PSNR}_{ori}}\times100\%,
	\end{equation}
	where ${Bitrate}_{ori}$ and ${PSNR}_{ori}$ represent the bit rate and PSNR value of direct JPEG compressed image, respectively, ${Bitrate}_{jnd}$ and ${PSNR}_{jnd}$ represent the bit rate and PSNR value of JND-guided JPEG compressed image, respectively. According this definition, it is reasonable to say a higher value of $G$ indicates a better performance of a JND model for guiding the JPEG compression. We thus evaluate the performance of JND-guided JPEG compression using $G$ and show the results in Table \ref{tab_compression}. As shown, our proposed JND model-guided JPEG compression is able to save $16.54\%$ of bitrates in average while Wu2017's JND model-guided JPEG compression can save $42.88\%$ of bitrates in average. However, our proposed JND model-guided JPEG compression only reduces $3.87\%$ of PSNR value in average while Wu2017's JND model-guided JPEG compression will reduce $14.90\%$ of PSNR value in average. By taking PSNR reduction and bitrate saving into account simultaneously with a single criteria $G$, our proposed JND model-guided JPEG compression achieves much larger values of $G$ for the majority images (except ``I11'') and also a much larger average value of $G$ when considering all the images together. 
	
	We further set different QP values and draw the curves of the average value of $G$ with different QP values, as depicted in Fig. \ref{fig_gain}. Our proposed JND model-guided JPEG compression consistently owns higher average value of $G$ than that of Wu2017's by a large margin at each QP value. This further demonstrates the superiority of our proposed JND model for improving the efficiency of JPEG compression. Based on these experimental results, we can savely conclude that: 1) in comparison with the direct JPEG compression, the JPEG compression guided by our proposed JND model could further save moderate bitrates while without reducing the visual quality too much; 2) in comparison with Wu2017's JND model, JPEG compression guided by our proposed JND model could achieve a better balance between visual quality reduction and bit rate saving.
	
	\begin{table}[!t]
		\centering
		\caption{RMSE between the distortion visibility maps predicted by VDP metrics and ground-truth maps marked by humans. JND2VDP represents the metric which is converted from our proposed JND model according to Eq. (19).}
		\label{tab_JND2VDP}
		\newcommand{\tabincell}[2]{\begin{tabular}{@{}#1@{}}#2\end{tabular}}
		\setlength{\tabcolsep}{5pt}
		\renewcommand\arraystretch{1.5}
		\resizebox{250pt}{!}{\begin{tabular}{c|c|c|c|c|c}
				\hline \hline
				\multirow{1}{*}{Subset} & HDRVDP 2.0 & HDRVDP 2.1 & HDRVDP 2.2.1 & HDRVDP 3.0.6 & JND2VDP \\  \hline \hline
				aliasing & 0.4450  & \textbf{0.2012}  & 0.2046  & \textbf{0.1315}  & 0.2344    \\  \hline
				cgibr & 0.7790  & \textbf{0.3714}  & 0.3773  & \textbf{0.1706}  & 0.4467     \\  \hline
				compression & 0.6958  & 0.5966  & 0.5950  & \textbf{0.3442}  & \textbf{0.5476}     \\  \hline
				deghosting & 0.5912  & \textbf{0.3488}  & 0.3495  & \textbf{0.3339}  & 0.6090     \\  \hline
				downsampling & 0.1524  & \textbf{0.0846}  & 0.0849  & \textbf{0.0732}  & 0.0861   \\  \hline
				ibr & 0.7415  & \textbf{0.2957}  & 0.3010  & \textbf{0.1211}  & 0.4130     \\  \hline
				mixed & 0.4495  &	\textbf{0.2243}  & 0.2289  & \textbf{0.1236}  & 0.2638   \\  \hline
				perceptionpatterns & 0.6916  & 0.4033  & 0.4068  & \textbf{0.2952}  & \textbf{0.3430}   \\ \hline
				peterpanning & 0.4643  & 0.3717  & 0.3746  & \textbf{0.1627}  & \textbf{0.1856}  \\  \hline 
				shadowacne & 0.4787  & 0.3316  & 0.3347  & \textbf{0.1630}  & \textbf{0.2011}    \\  \hline
				tid2013 & 0.5060  & \textbf{0.4380} 	& \textbf{0.4400}  & 0.4760	& 0.5038 	   \\  \hline
				zfighting & 0.4664  & 0.3685 	& 0.3725  & \textbf{0.1290}	& \textbf{0.1963}   	 \\  \hline
				average & 0.5384  &	0.3363  & 0.3391  & \textbf{0.2103}  & \textbf{0.3359}   \\  \hline	\hline	
		\end{tabular}}
	\end{table}
	
	\begin{figure}[!t]
		\centering
		\includegraphics[width=\linewidth]{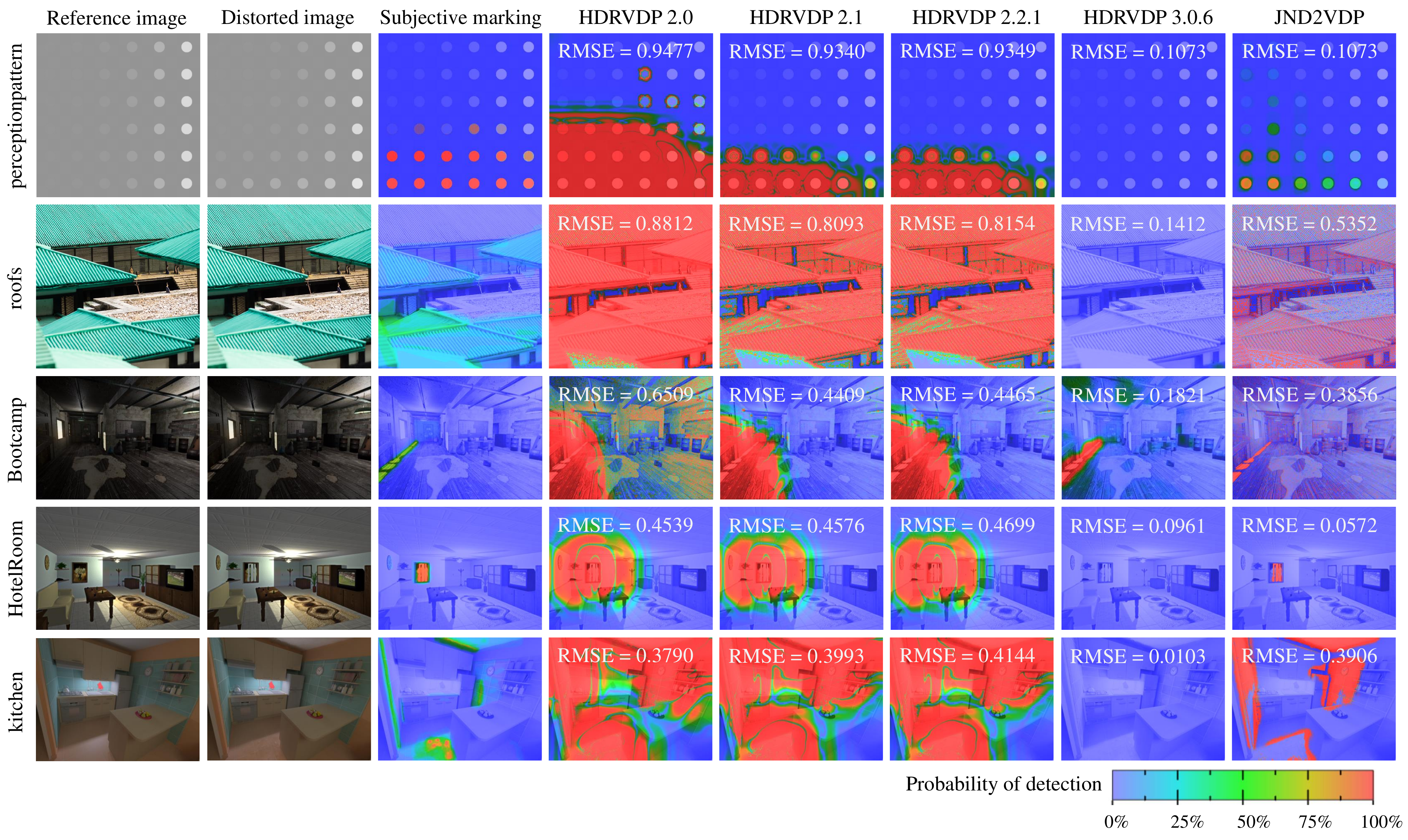}
		\caption{Visual comparisons of the predicted visibility map results. JND2VDP represents the metric which is converted from our proposed JND model according to Eq. (19).}
		\label{JND2VDP}
	\end{figure}

	\begin{table*}[!t]
		\centering
		\caption{Running Time Comparison.}
		\label{tab_runningtime}
		\newcommand{\tabincell}[2]{\begin{tabular}{@{}#1@{}}#2\end{tabular}}
		\setlength{\tabcolsep}{5pt}
		\renewcommand\arraystretch{1.3}
		\resizebox{500pt}{!}{\begin{tabular}{c|c|c|c|c|c|c|c|c}
				\hline \hline
				\multirow{1}{*}{Models} & Yang2005 \cite{Yang2005} & Zhang2005 \cite{Zhang2005} & Wu2013 \cite{Wu2013} & Wu2017 \cite{Wu2017} 
				& Jakhetiya2018 \cite{JAKHETIYA} & Chen2020 \cite{Chen2020} & Shen2021 \cite{Shen2021} & Proposed  \\  \hline 
				Time (s) &$\approx$0.1372&	$\approx$5.8350&	$\approx$9.1736&	$\approx$0.9706	&$\approx$57.5209&	$\approx$0.6843&	$\approx$9.8281&	$\approx$0.2941   \\  \hline \hline	
		\end{tabular}}
	\end{table*}
	
	\subsection{Comparison with VDP-type Metrics}
	As stated in Eqs. (1) and (2), the JND and VDP are two relevant concepts in that they both attempt to model the same underlying mechanism of the visual system - detection and discrimination. To enable a direct comparison between the JND model and VDP-type metrics, we need to convert the JND map into an approximate visibility map in the following manner:
	\begin{equation}
		O_{JND2VDP} \approx p\left( (I_d-I_o)/F_{JND}(I_o) \right),
	\end{equation}
	where $p(x)$ is the psychometric function. According to Eq. (17) in \cite{HDR-VDP2}, $p(x)$ can be expressed as follows:
	\begin{equation}
		p(x) = 1-\text{exp}(\text{log}(0.5)x^{\beta}),
	\end{equation}
	where $\beta$ is the slope of the psychometric function and the value set to $\beta = 3.5$ \cite{VDP}.
	
	The JND-converted visibility map is denoted by JND2VDP in the following. In the experiments, we adopt the distortion visibility dataset provided in \cite{2018TOG} as the benchmark. The JND2VDP is compared with HDR-VDP 2.0 \cite{HDR-VDP2}, HDR-VDP 2.1 \cite{HDR-VDP2}, HDR-VDP 2.2.1 \cite{HDR-VDP2,HDR-VDP2.2}, and HDR-VDP 3.0.6 \cite{HDR-VDP2}. The details of comparison are illustrated as follows. For each scene in the dataset, there is a reference image, a distorted image, and a subjective marking map. The subjective marking map reflects the area and intensity of visible distortions in the distorted image. A larger intensity value in the subjective marking map means a larger probability of detecting the distortion in the distorted image. The values in the subjective marking map are normalized in the range [0,1] for computation. We take the subjective marking map as the ground-truth, and we compute the RMSE value between subjective marking map and the probability map produced by JND2VDP or other compared VDP metrics. A smaller RMSE value means a better prediction accuracy. The comparison results are shown in Table \ref{tab_JND2VDP}. From this table, we can observe that JND2VDP is only worse than the latest HDR-VDP 3.0.6 while slightly better than HDRVDP 2.0 \cite{HDR-VDP2}, HDRVDP 2.1 \cite{HDR-VDP2}, and HDRVDP 2.2.1 \cite{HDR-VDP2,HDR-VDP2.2}. In Fig. \ref{JND2VDP}, we further visualize some results of the predicted visibility maps for reference. These results demonstrate that our predicted JND map has a good capability in distortion detection and discrimination. 
	
	\subsection{Running Time}
	Besides the high prediction accuracy, an excellent JND model should also be computationally efficient. We test the running time of different JND models with the same setting and platform. The experiments are all conducted on a PC with an AMD Ryzen 7 4800H @ 2.9 GHZ and 16GB RAM. The software platform is MATLAB R2019a. The running times of different JND models are compared in Table \ref{tab_runningtime}. We present the running times in the unit of second (s) and a smaller value means a faster running speed. Our proposed JND model ranks the second place among all competitors with a running speed within 0.3ms for processing a 1200$\times$800 image. Although Yang2005 is more efficient, the inferior prediction accuracy makes it unsuitable for using in practical applications.
	
	\subsection{Limitation}
	Although the proposed JND model can achieve better performance than the existing ones, it also has limitations. Our model is also based upon data obtained in one subjective experiment and its parameters are fully dependent on the experimental conditions of the subjective test. Thus, our JND model is only suitable for that viewing condition of test and cannot flexibility adapt for different viewing conditions. According to \cite{viewing-condition1,viewing-condition2}, it is known that viewing conditions will have direct influence on the visually lossless threshold of images. The user studies in \cite{viewing-condition1,viewing-condition2} also show that higher peak luminance and shorter viewing distance will help users to more easily discover distortions. However, our work does not take into account the influence of viewing condition, i.e., we only focus on predicting the JND of images under a specific viewing condition setting. In other words, our work only accounts for image content while ignoring the influence of viewing condition. For example, the standard viewing conditions are mostly at 60 ppd while we set ppd=30 as an alternative to setup our subjective experiments. We admit that such an ignorance of viewing condition is an obvious limitation of our method. In the future, we will devote to considering viewing condition as an influential factor toward building a more flexible and effective JND prediction models in practical applications.
	
	Another limitation is that the concept of a JND map does not account for the notion of energy or spatial pooling. For example, if we change a single pixel by the value from the JND map, the change is unlikely to be detected. But, if we change a large number of pixels (according to the JND map), the change is likely to be well-visible. The type of change will also have a strong effect on the visibility of changes. If we add the same value to all pixels (e.g., the minimum from the JND map), the change is unlikely to be detected because we are insensitive to low-frequency or direct component (DC) brightness changes. But, if we add salt-and-paper noise of the same amplitude across the image, the change is likely to be observed. The JND maps cannot distinguish between those cases.
	
	\section{Conclusion}
	This paper has presented a novel top-down JND estimation model of natural images. It dedicates to estimating a CPL image first by exploiting the distribution characteristic of the cumulative normalized KLT coefficient energy and then calculating the difference map bettween the original image and CPL image as the JND map. The difference map well reflects the redundant micro-structure information which typically cannot be perceived by the HVS. Using such a difference map as the final JND map, the visual redundancies in the image can be better exploited. We have evaluated the performance of the proposed JND model explicitly with direct JND prediction and implicitly with two applications including JND-guided noise injection and JND-guided image compression. Experimental results have demonstrated that our proposed JND model can achieve better performance than several latest JND models. In addition, we also compare the proposed JND model with existing VDP metrics in terms of the capability in distortion detection and discrimination. The results indicate that our proposed JND model also has a good performance in this task.
	
	
	\small
	\bibliographystyle{IEEEtran}
	\bibliography{Bibliography-File}

	\begin{IEEEbiography}[{\includegraphics[width=1in,height=1.33in,clip]{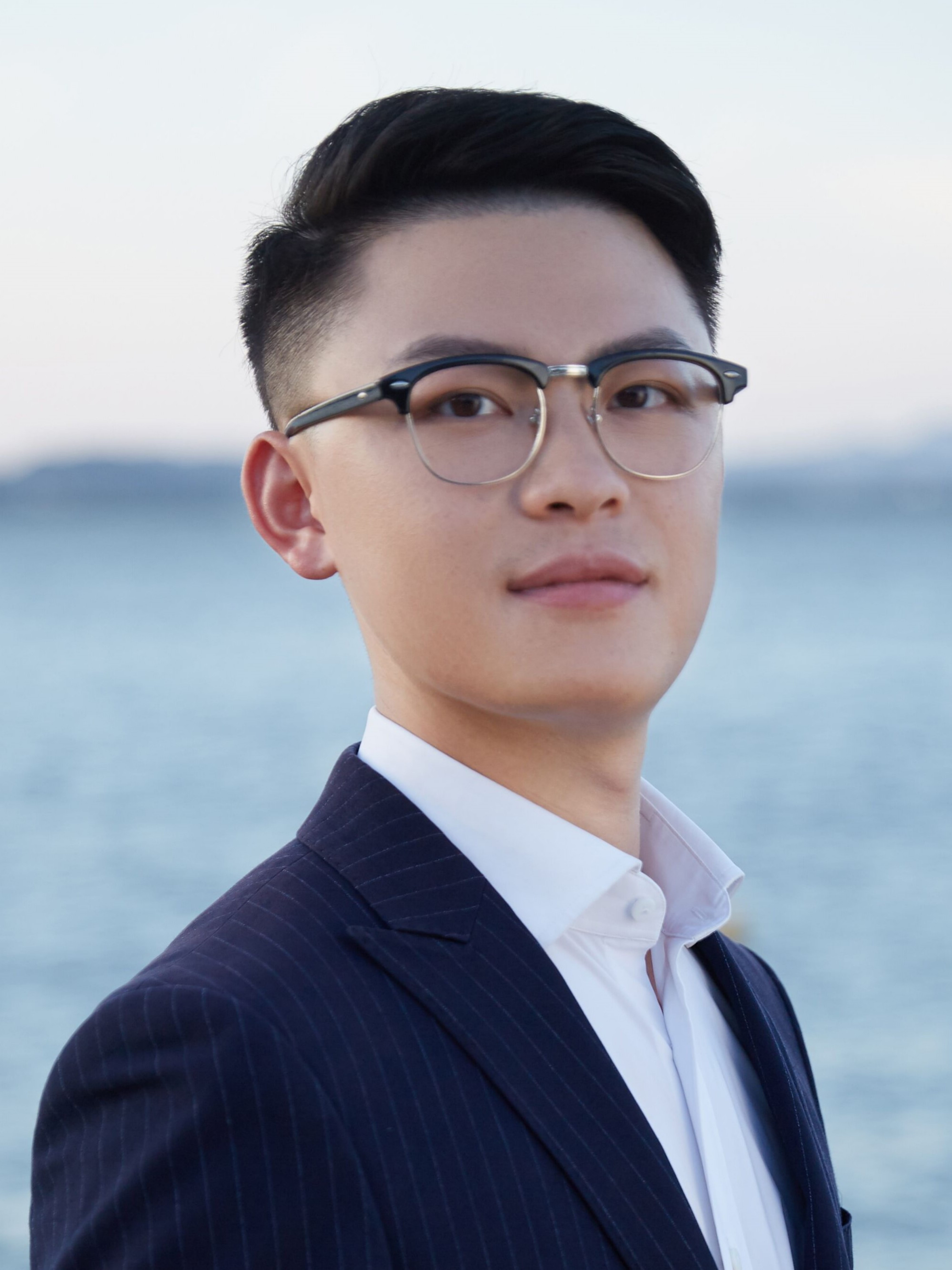}}]{Qiuping Jiang} is currently an Associate Professor with Ningbo University, Ningbo, China. He received the Ph.D. degree in Signal and Information Processing from Ningbo University in 2018. From Jan. 2017 to May 2018, he was a visiting student with Nanyang Technological University, Singapore. His research interests include image processing, visual perception modelling, and deep learning with applications in computer vision. He received the Best Paper Honorable Mention Award of the Journal of Visual Communication and Image Representation. He also serves as the Associate Editor for IET Image Processing, Journal of Electronic Imaging, and APSIPA Trans. on Information and Signal Processing.
	\end{IEEEbiography}

	\vspace{8ex}
	
	\begin{IEEEbiography}[{\includegraphics[width=1in,height=1.33in,clip]{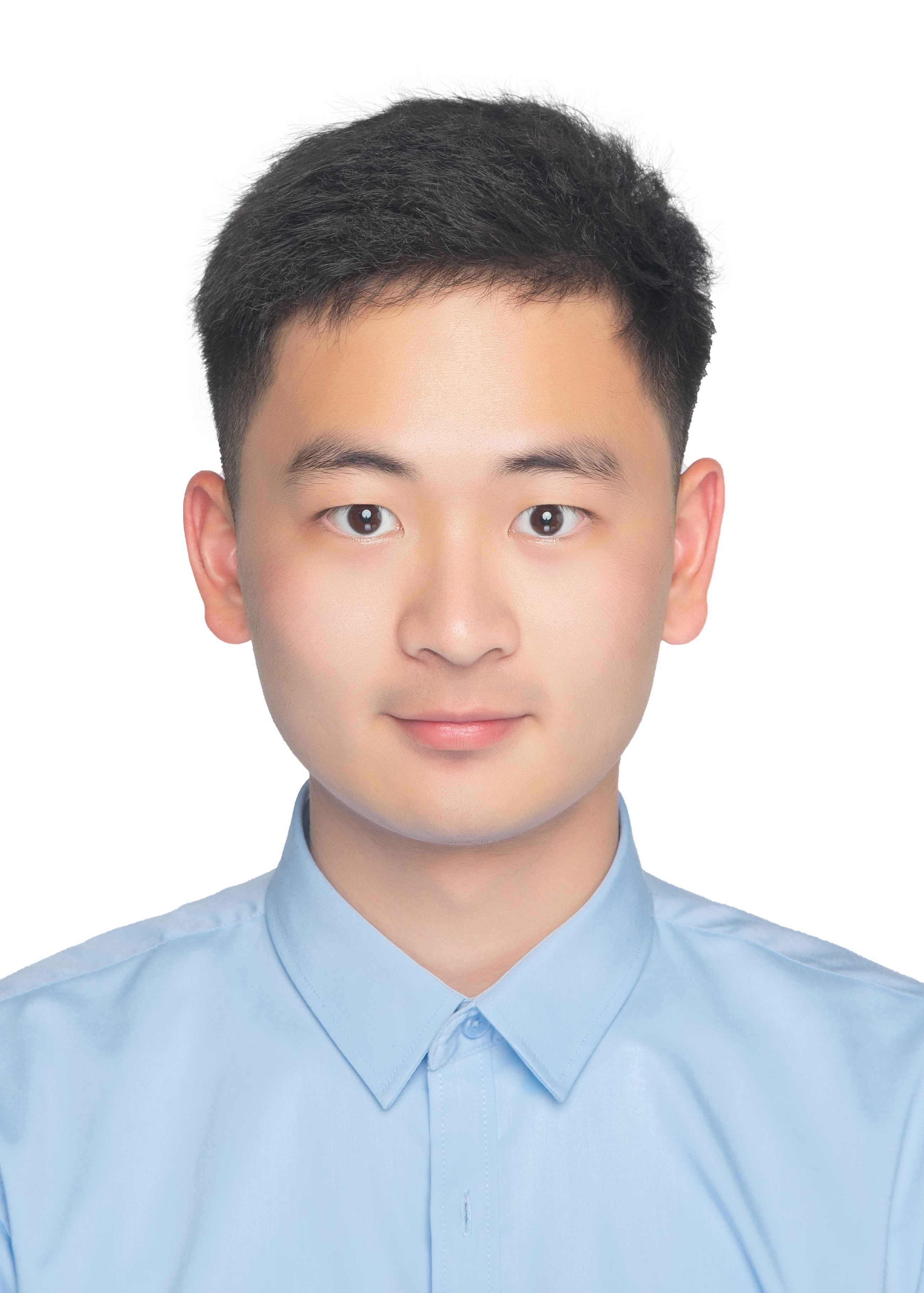}}]{Zhentao Liu} is a final year undergraduate student major in Communication Engineering with the School of Information Science and Engineering, Ningbo University, China. His research interests include image processing, image quality assessment, and visual perception modeling. He will continue pursuing the master's degree in Computer Science and Technology at ShanghaiTech University.
	\end{IEEEbiography}		
	
	\begin{IEEEbiography}[{\includegraphics[width=1in,height=1.25in,clip]{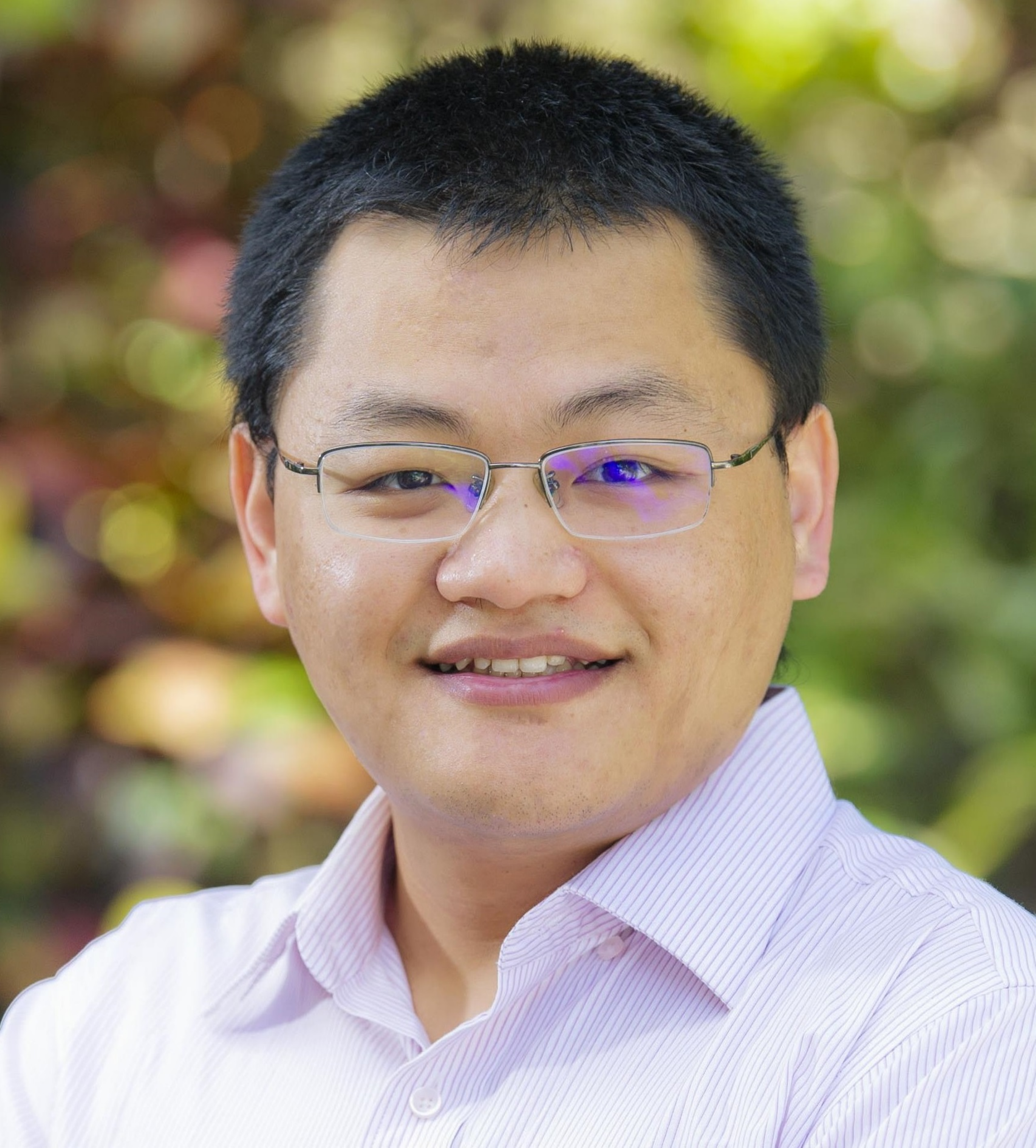}}]{Shiqi Wang} received the B.S. degree in computer science from the Harbin Institute of Technology in 2008, and the Ph.D. degree in Computer Application Technology from the Peking University under the supervision of Prof. Wen Gao, in 2014. From Mar. 2014 to Mar. 2016, He was a Postdoc Fellow with the Department of Electrical and Computer Engineering, University of Waterloo, Waterloo, Canada. From Apr. 2016 to Apr. 2017, He was with the Rapid-Rich Object Search Laboratory, Nanyang Technological University, Singapore, as a Research Fellow. He is currently an Assistant Professor with the Department of Computer Science, City University of Hong Kong. His research interests include video compression, image/video quality assessment, and image/video search and analysis. He received the Best Paper Award in IEEE VCIP 2019, ICME 2019, IEEE Multimedia 2018, and PCM 2017. His coauthored article received the Best Student Paper Award in IEEE ICIP 2018. He serves as an Associate Editor for IEEE Transactions on Circuits and Systems for Video Technology.
	\end{IEEEbiography}
	
	\begin{IEEEbiography}[{\includegraphics[width=1in,height=1.2in,clip]{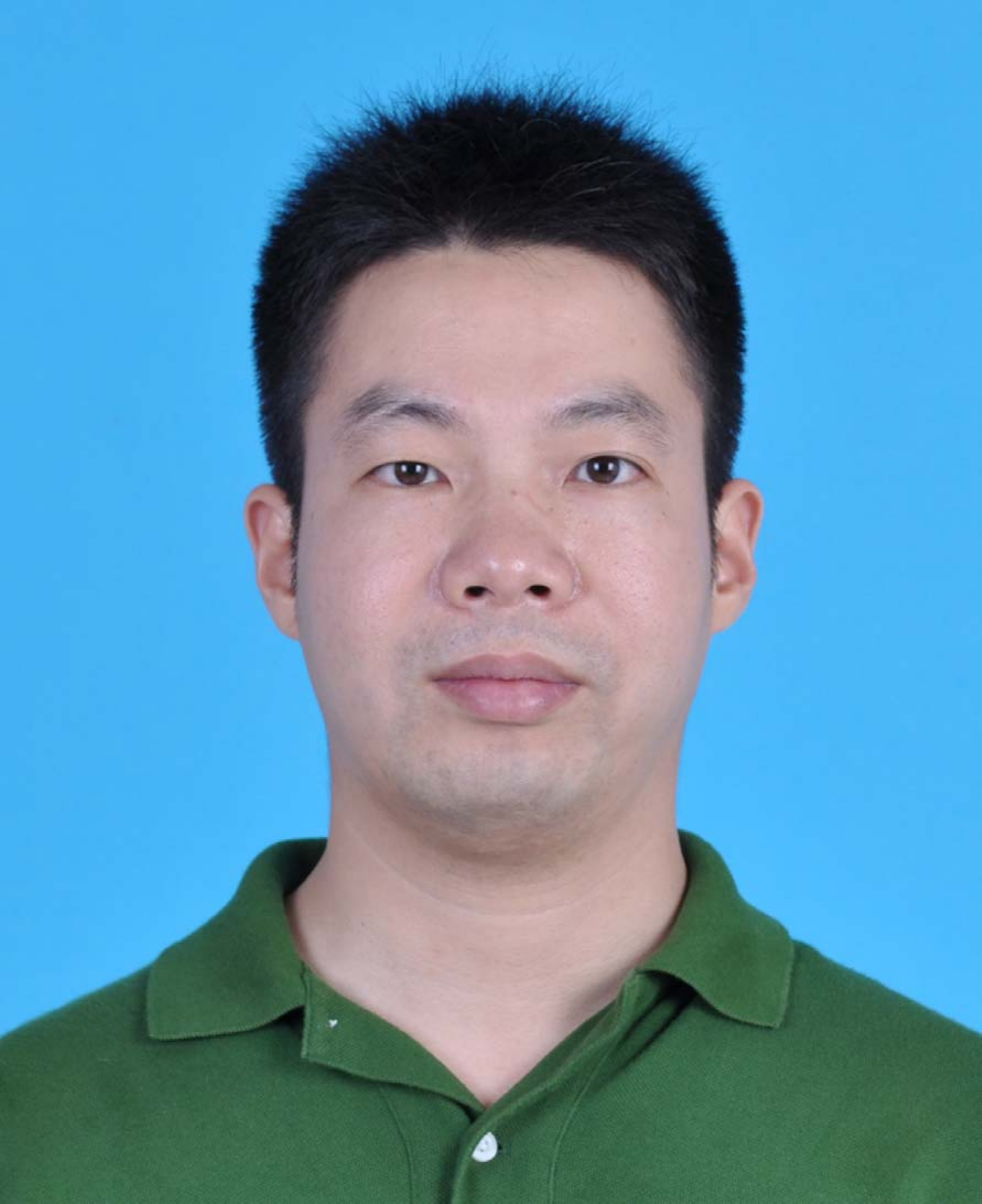}}]{Feng Shao} received the B.S. and Ph.D. degrees in Electronic Science and Technology from Zhejiang University, Hangzhou, China, in 2002 and 2007, respectively. He is currently a Professor with the School of Information Science and Engineering, Ningbo University, Ningbo, China. His research interests include image processing, image quality assessment, and immersive media computing.
	\end{IEEEbiography}
	
	\begin{IEEEbiography}[{\includegraphics[width=1in,height=1.2in,clip]{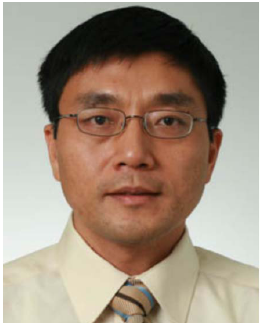}}]{Weisi Lin} is currently a Professor with the School of Computer Science and Engineering, Nanyang Technological University, Singapore. He received the Bachelor's degree in Electronics and then a Master's degree in Digital Signal Processing from Sun Yat-Sen University, Guangzhou, China, and the Ph.D. degree in Computer Vision from King’s College, London University, UK. His research interests include image processing, perceptual modeling, video compression, multimedia communication, and computer vision. \par
	He is a Fellow of the IEEE and IET, an Honorary Fellow of the Singapore Institute of Engineering Technologists, and a Chartered Engineer in U.K. He was the Chair of the IEEE MMTC Special Interest Group on Quality of Experience. He was awarded as the Distinguished Lecturer for IEEE Circuits and Systems Society in 2016-2017. He served as a Lead Guest Editor for a Special Issue on Perceptual Signal Processing of the IEEE JOURNAL OF SELECTED TOPICS IN SIGNAL PROCESSING in 2012. He also has served or serves as an Associate Editor for IEEE Transactions on Image Processing, IEEE Transactions on Circuits and Systems for Video Technology, IEEE Transactions on Multimedia, IEEE Signal Processing Letters, and Journal of Visual Communication and Image Representation.
	\end{IEEEbiography}

\end{document}